\documentclass[preprint,11pt,doublespace,showkeys,preprintnumbers,prl]{revtex4-1}

\usepackage[latin9]{inputenc}
\usepackage{xcolor}
\usepackage{float}
\usepackage{amsmath}
\usepackage{amssymb}
\usepackage{amsfonts}
\usepackage{graphicx}
\usepackage{esint}

\makeatletter
\definecolor{GREEN}{rgb}{0,1,0}

\usepackage{mathrsfs}
\usepackage{color}

\usepackage{ulem}

\makeatother

\begin{document}

\title{
Nonlinear phase-amplitude reduction of delay-induced oscillations
}

\author{Kiyoshi Kotani
\footnote{Electric adress: kotani@neuron.t.u-tokyo.ac.jp}
}
\affiliation{
Research Center for Advanced Science and Technology, The University of Tokyo, 4--6--1 Komaba, Meguro-ku, Tokyo 153-8904, Japan\\
and JST, PRESTO, 4--1--8 Honcho, Kawaguchi-shi, Saitama 332-0012, Japan
}

\author{Yutaro Ogawa }
\affiliation{
Research Center for Advanced Science and Technology, The University of Tokyo, 4--6--1 Komaba, Meguro-ku, Tokyo 153-8904, Japan
}

\author{Sho Shirasaka}
\affiliation{
Graduate School of Information Science and Technology, Osaka University,
1-5 Yamadaoka, Suita, Osaka 565-0871, Japan\\
and Research Center for Advanced Science and Technology, The University of Tokyo, 4--6--1 Komaba, Meguro-ku, Tokyo 153-8904, Japan 
}

\author{Akihiko Akao}
\affiliation{
Department of Mathematics, University of Pittsburgh,
Pittsburgh, Pennsylvania 15213, USA\\
and Graduate School of Engineering, The University of Tokyo, 7--3--1 Hongo, Bunkyo-ku, Tokyo 113-0033, Japan
}

\author{Yasuhiko Jimbo}
\affiliation{
Graduate School of Engineering, The University of Tokyo, 7--3--1 Hongo, Bunkyo-ku, Tokyo 113-0033, Japan
}

\author{Hiroya Nakao}
\affiliation{Department of Systems and Control Engineering, Tokyo Institute of Technology, Tokyo 152-8522, Japan
}



\begin{abstract}
Spontaneous oscillations induced by time delays are observed in many real-world systems.
Phase reduction theory for limit-cycle oscillators described by delay-differential equations (DDEs) has been developed to analyze their synchronization properties, but it is applicable only when the perturbation applied to the oscillator is sufficiently weak.
In this study, we formulate a nonlinear phase-amplitude reduction theory for limit-cycle oscillators described by DDEs on the basis of the Floquet theorem, which is applicable when the oscillator is subjected to perturbations of moderate intensity.
We propose a numerical method to evaluate the leading Floquet eigenvalues, eigenfunctions,
and adjoint eigenfunctions necessary for the reduction and derive a set of low-dimensional
nonlinear phase-amplitude equations approximately describing the oscillator dynamics.
By analyzing an analytically tractable oscillator model with a cubic nonlinearity,
we show that the asymptotic phase of the oscillator state in an infinite-dimensional state space can be approximately
evaluated and non-trivial bistability of the oscillation amplitude
caused by moderately strong periodic perturbations can be predicted on the basis of the derived phase-amplitude equations.
We further analyze a model of gene-regulatory oscillator and illustrate that the reduced equations can elucidate the mechanism of its complex dynamics under  non-weak perturbations, which may be relevant to real physiological phenomena such as circadian rhythm sleep disorders.
\end{abstract}

\pacs{05.45.Xt, 02.30.Ks}

\keywords{time-delayed systems \textbar{} Floquet theorem \textbar{}  limit-cycle oscillations \textbar{} synchronization  \textbar{} adjoint method \textbar{} dimensionality reduction \textbar{} multistability}

\date{\today}
\maketitle

\section{Introduction \label{sec. intro}}

Time-delayed feedback can break continuous time-translational symmetry and lead to oscillatory behavior in many physical, biological, social, and engineered systems~\cite{Glass+,Lewis+, Dfiremother, Brent+, Peterka+,Soriano,Kalmar+,Szydlowski+}.
In biology, for example, ultradian oscillations in the hypothalamic-pituitary-adrenal (HPA) system are induced by time-delayed synthesis of hormones in the adrenal cortex~\cite{Walker+}.
Also, somite segmentation in zebrafish is regulated by oscillatory dynamics induced by time delays in the synthesis of proteins~\cite{Ishimatsu}, and mammalian circadian rhythm is generated by feedback regulations of clock genes in suprachiasmatic nucleus (SCN)~\cite{Lema+}.
Such oscillatory dynamics in systems with time delays can be described as stable limit-cycle orbits of delay-differential equations (DDEs).

In many of such systems, each oscillatory unit, or oscillator, is not isolated but perturbed by external forcing or by mutual coupling with other oscillators, and the state of each oscillator may deviate from the unperturbed limit cycle of an isolated oscillator when the perturbation is not sufficiently weak.
Therefore, it is important to understand how perturbations of moderate intensity can modulate the period, amplitude, and other properties of delay-induced oscillations.

For example, in the case of zebrafish somite segmentation, it is known that strong couplings between cells are necessary for the spatio-temporal oscillatory dynamics of {\it her1} (zebrafish {\it hairy-related gene1}) expression~\cite{Ishimatsu}.
In the case of circadian clock genes, the oscillatory period in the free-running condition is known to be slightly different from 24 h, but they are entrained by the periodic external day-and-night lights through retinal ganglion cells~\cite{SackI, Berson}.
Strong light stimulation can further induce large modulation in the activities of the clock genes~\cite{Ukai+}. 
Since irregular dynamics of circadian rhythms manifest themselves as diseases such as sleep disorders~\cite{SackI,  Thorpy+, Jones, Moldofsky}, understanding of the dynamics of circadian clock genes under strong perturbations may facilitate therapies for sleep disorders.

The phase reduction theory is a standard mathematical framework for characterizing response properties of weakly perturbed limit-cycle oscillators and analyzing their synchronization dynamics via dimensionality reduction~\cite{Winfree2,Kuramoto,Pikovsky,Ermentrout2,Nakao, KuramotoPTRSA, Ashwin}.
Recently, the phase reduction theory has been extended also to DDEs exhibiting stable limit-cycle oscillations,
which requires non-trivial mathematical generalization because DDEs are infinite-dimensional dynamical systems~\cite{Kotani,Pyragas}.
However, the phase reduction has a strong limitation in that it is applicable only when the oscillator state remains sufficiently close to the unperturbed limit cycle.
Specifically, when non-weak perturbations are applied or relaxation time of the system state to the limit cycle is not sufficiently small, the amplitude degrees of freedom may no longer be enslaved by the phase, leading to the breakdown of the lowest-order phase-only description. In such cases, the nonlinear interaction of the phase and amplitude may lead to non-trivial dynamics that cannot be captured by the phase reduction.

To overcome this difficulty, several mathematical frameworks
have been proposed for oscillatory systems described by ordinary differential equations (ODEs), such as
higher-order phase resetting curves~\cite{Canavier+}, extended phase equations~\cite{Rubin+}, and higher-order phase-amplitude equations~\cite{Castejon+}.
Still, for oscillatory dynamics of DDEs away from the limit cycle, much remains unknown because of their infinite-dimensional nature.
Thus, a general framework for dimensionality reduction of limit-cycle oscillators described by DDEs that can analyze the effect of moderately strong perturbations is needed. Such a framework would shed light on oscillatory dynamical systems in which nonlinearity, time delay, and strong perturbations coexist.

In this study, our interest lies in the situation where the phase and amplitude of DDEs interact significantly in a nonlinear manner. 
We develop a nonlinear phase-amplitude reduction theory for DDEs, which gives a general mathematical framework for reducing DDEs describing limit-cycle oscillators to low-dimensional ordinary differential equations on the basis
of the Floquet theory~\cite{Stokes,Hale,Simmendinger}.
We also propose a practical numerical method, which we call the extended adjoint method, to evaluate the Floquet eigenvalues, eigenfunctions, and their adjoint functions, which are necessary for the reduction.
By using biorthogonality of the Floquet eigenfunctions and their adjoints, we project
the oscillator state onto an eigenspace spanned by a few slowly-decaying Floquet eigenfunctions and derive a set of phase-amplitude equations which takes into account nonlinear interactions between the slowly-decaying Floquet eigenmodes.
In contrast to the standard lowest-order phase reduction, the amplitude component associated with
the second Floquet eigenfunction is included, which can play important roles when the relaxation of the system is slow or when the system is strongly perturbed.

We confirm the validity of the theory using an analytically-tractable DDE
with a cubic nonlinearity by showing that the reduced phase-amplitude equations
accurately predict the amplitude of the phase-locked oscillations under a periodic force, which exhibits non-trivial bistable 
response induced by the non-weak amplitude effects.
We then apply the theory to a model of a gene-regulatory oscillator under 
moderately strong forcing and analyze its synchronization dynamics. 
We show that the reduced phase-amplitude equations can also predict
nontrivial bistable dynamics of the system, which is analogous to a circadian disorder
called advanced sleep-phase syndrome (ASPS)~\cite{Thorpy+, Jones, Moldofsky}.

\section{Theory}

In this section, we derive a set of reduced nonlinear phase-amplitude equations for limit-cycle oscillators described by DDEs on the basis of the Floquet theory and propose a practical numerical method to calculate the Floquet eigenvalues, eigenfunctions and their adjoints that are necessary for the reduction. We also derive approximate phase-amplitude equations for the oscillators subjected to periodic external forcing.

\subsection{DDEs with a stable limit-cycle solution}

We consider general delay-differential equations (DDEs) that have a stable limit-cycle solution.
Mathematical analysis of such DDEs, for example, analyzing the synchronization properties when they are periodically perturbed, is not easy because they describe nonlinear infinite-dimensional dynamical systems on Banach spaces.
Our aim is to derive simpler tractable equations by reducing them to low-dimensional ODEs while preserving their essential quantitative properties and to analyze synchronization dynamics of nonlinear oscillators described by such DDEs under moderately strong external perturbations.
In previous studies~\cite{Kotani,Pyragas}, phase reduction methods for stable limit-cycle solutions of DDEs have been developed, which are applicable when the perturbations given to the system is sufficiently weak.
In this study, we develop a nonlinear phase-amplitude reduction theory for DDEs.

We consider a DDE for $X(t)\in \mathbb{R}^{N}$, represented as a column vector, with a maximum delay time $\tau > 0$.
To construct a solution of the DDE, we have to take into account the history of $X(t)$ from $t-\tau$ to $t$.
Thus, we introduce its history-function representation, 
$X^{(t)}({\sigma})\equiv X(t+{\sigma})$ $(-\tau\leq\sigma\leq0)$~\cite{Stokes,Hale,Simmendinger}.
Here, $X^{(t)}(\cdot)\in C_0$ and $C_{0}=C([-\tau,0]\to \mathbb{R}^{N})$ is a Banach space of (column) vector-valued continuous functions mapping $[-\tau , 0]$ into $\mathbb{R}^N$, which is equipped with a norm $||x||_{C_0} = \sup_{\theta \in [-\tau,0]}||x(\theta)||$, where $||\cdot||$ is the usual Euclidean norm on $\mathbb{R}^{N}$.
This history function $X^{(t)}$ represents the state of the dynamical system described by the DDE at time $t$, where the state space of the system is given by the infinite-dimensional Banach space $C_0$. 

Using the above notation, a DDE can generally be written as
\begin{eqnarray}
{\frac{d}{dt}}X^{(t)}(\sigma)=
\left\{ \begin{array}{lr}
{\displaystyle {\displaystyle {\frac{d}{d\sigma}}X^{(t)}(\sigma)}} & (-\tau\leq\sigma<0),\\
\\
{\displaystyle \mathcal{N}(X^{(t)} (\cdot) )}+G\left(X^{(t)} (\cdot), t\right) & (\sigma=0).
\end{array}\right.\label{2}
\end{eqnarray}
Here, the vector-valued functional $\mathcal{N} : C_0 \to {\mathbb R}^N$ represents the system dynamics
and $G: C_0 \times \mathbb{R} \to {\mathbb R}^N$ denotes external perturbation applied to the system that depends on the system state $X^{(t)}$. Both functionals are assumed to be sufficiently smooth.
This DDE can describe not only systems with discrete delays but also systems with distributed delays~\cite{Wischert}; see Ref.~\cite{Palm} for the relation between nonlinear functionals and their kernel representations, which are widely used for systems with distributed delays
described by integro-differential equations.

We consider a situation in which the DDE~(\ref{2}) without the external perturbation ($G=0$) has a stable limit-cycle solution $X_{0}(t)$ whose period is $T$, i.e., $X_{0}(t+T) = X_{0}(t)$, and represent it as a history function $X_0^{(t)} (\cdot) \in C_0$
satisfying $X_0^{(t+T)} = X_0^{(t)}$, where
\begin{align}
X_0^{(t)}(\sigma) \equiv X_0(t+\sigma) \quad (-\tau\leq\sigma\leq0).
\end{align}
In what follows, we also denote the limit cycle as $X_{0}^{(\phi)}$, where we use the phase $\phi$ ($0 \leq \phi < T$) in place of the time $t$ to parametrize system state on the limit cycle. The phase $\phi$ increases from $0$ to $T$, where the origin $\phi = 0$ can be chosen as a specific system state on the limit cycle. When the system state evolves along the limit cycle without perturbation, the phase $\phi$ increases with a constant frequency $1$, i.e., $\phi = t \ (\mbox{mod}\ T)$.
Similarly, we also denote $T$-periodic history functions, such as the Floquet eigenfunctions, using the phase $\phi$ instead of $t$ when necessary. 

The definition of the phase can further be extended to the basin of attraction of the limit cycle by assigning the same phase value $\phi$ to the set of system states $\{ X^{(t)} \}$ that asymptotically converge to the same system state as $X_{0}^{(\phi)}$ when the system evolves without perturbation~\cite{Kotani,Pyragas}, i.e., 
$\lim_{t \to \infty} \| X^{(t)} - X_0^{(\phi + t)} \|_{C_0}= 0$, yielding the notion of {\it asymptotic phase} $\Phi(X^{(t)}) : C_0 \to [0, T)$ that maps a system state $X^{(t)}$ in the basin to a phase value.
The asymptotic phase $\Phi$ satisfies
\begin{equation}
\frac{d}{dt} \Phi(X^{(t)})=1
\label{eq. as_p}
\end{equation}
when the system state evolves in the basin of the limit cycle without perturbation.
The isosurfaces of $\Phi$, called {\it isochrons}, are not simply hyperplanes in general. 
For ordinary differential equations, the asymptotic phase has been used as a canonical representation of rhythms of stable oscillatory dynamics~\cite{Winfree2,Kuramoto,Pikovsky,Ermentrout2,Hale_ODE} and provides in-depth insights into strongly-perturbed oscillatory dynamics~\cite{Ashwin,Canavier+,Rubin+,Castejon+,Rabinovitch+}. Recently, it has also been defined for DDEs and other non-conventional oscillatory systems~\cite{Kotani,Pyragas}.

We assume that the relaxation dynamics of the system state to the limit cycle can be decomposed into a few slow modes and remaining faster modes, which are well separated in time scale from each other.
In this case, a rectangular coordinate frame moving along the periodic orbit, which was used in Refs.~\cite{Hale_ODE,Wedgwood,Medvedev}, is not useful for reducing the dynamics to low-dimensional ODEs, because fast and slow components interact already at the lowest order in this coordinate frame. 
It is also not easy to proceed with the asymptotic phase and associated amplitudes, because they are generally given by highly nonlinear functionals of the system state $X{(t)}$.
We therefore use a coordinate frame defined by the Floquet eigenfunctions to decompose the system state as discussed in Ref.~\cite{KuramotoPTRSA} for ODEs. 
The space spanned by the Floquet eigenfunctions with non-vanishing relaxation rates is tangent to the isochron at each point on the limit cycle.
For this purpose, we need to calculate the Floquet eigenvalues, eigenfunctions, and their adjoints of DDEs.

\subsection{Floquet theory for DDEs}

We first describe the Floquet theory for the DDE~(\ref{2}) without 
the perturbation term, i.e., $G = 0$.
We denote small deviation of $X(t)$ from $X_{0}(t)$ as $Y(t)=X(t)-X_{0}(t)$, and introduce its history-function representation
$Y^{(t)}(\cdot) \in C_{0}$ with $Y^{(t)}({\sigma}) \equiv Y(t+{\sigma})$ $(-\tau\leq\sigma\leq0)$ as
\begin{align}
Y^{(t)}({\sigma}) = X^{(t)}({\sigma}) - X_0^{(t)}({\sigma}) \quad (-\tau\leq\sigma\leq0).
\end{align}
The linearized variational equation for $Y^{(t)}$ is given by
\begin{equation}
\frac{d}{dt}Y^{(t)}(\sigma)=L^{(t)}\left(Y^{(t)} \right) (\sigma) \quad (-\tau\leq\sigma\leq0), \label{eq. var}
\end{equation}
where  $L^{(t)}(Y^{(t)})$ is a history representation of a linear functional defined by 
\begin{eqnarray}
L^{(t)}\left(Y^{(t)}\right)(\sigma)=\left\{ \begin{array}{lr}
{\displaystyle \frac{d}{d\sigma}Y^{(t)}(\sigma)} & (-\tau\leq\sigma<0),\\
\\
{\displaystyle \int_{-\tau}^{0}d\sigma'{\bf \bar{\Omega}}^{(t)}(\sigma')Y^{(t)}(\sigma')} & (\sigma=0).
\end{array}\right.\label{2-1}
\end{eqnarray}
Here, 
\begin{align}
{\bf \bar{\Omega}}^{(t)}(\sigma) \equiv \left. \frac{\delta {\mathcal N}(X^{(t)}(\cdot))}{\delta X^{(t)}(\sigma)} \right|_{X^{(t)}=X_0^{(t)}}
\end{align}
is a functional differentiation of $\mathcal{N}$ with respect to $X^{(t)}$ evaluated at the system state $X^{(t)} =X_0^{(t)}$ on the limit cycle.
Note that Eq.~(\ref{eq. var}) gives a periodically driven linear system because $X_0^{(t)}$ is $T$-periodic.
In what follows, we expand $\mathcal{N}$ in a functional Taylor series in $Y^{(t)}$ as
\begin{align}
\mathcal{N}( X^{(t)} (\cdot) ) = \mathcal{N}( X_0^{(t)} (\cdot) ) + L^{(t)} \left( Y^{(t)} \right)(0) + F_{\mathrm{nl}}\left(Y^{(t)}(\cdot)\right),
\label{eq:taylorexpandN}
\end{align}
where $L^{(t)} \left( Y^{(t)} \right)(0)$ represents a linear functional of $Y^{(t)}$ defined in Eq.~(\ref{2-1}) with $\sigma = 0$
and $F_{\mathrm{nl}}(Y^{(t)}(\cdot))$ represents the remaining nonlinear functional of $Y^{(t)}$, respectively, and both of these functionals are evaluated at $X^{(t)} = X_0^{(t)}$.

As an example, let us consider a simple DDE,
\begin{equation}
\frac{d}{dt}X(t)=\mathcal{N}(X(t),X(t-\tau)), \label{eq. simpledde}
\end{equation}
which is equivalent to
\begin{eqnarray}
{\frac{d}{dt}}X^{(t)}(\sigma)=
\left\{ \begin{array}{lr}
{\displaystyle {\displaystyle {\frac{d}{d\sigma}}X^{(t)}(\sigma)}} & (-\tau\leq\sigma<0),\\
\\
{\displaystyle \mathcal{N}(X^{(t)}(0),X^{(t)}(-\tau))} & (\sigma=0),
\end{array}\right.\label{eq. simplesg}
\end{eqnarray}
in the history-function representation.
By using the chain rule for functional differentiation and representing the terms in ${\mathcal N}$ as
\begin{eqnarray}
X^{(t)}(0) &=& \int_{-\tau}^0{d\sigma'\delta (\sigma') X^{(t)}(\sigma')}
\end{eqnarray}
and
\begin{eqnarray}
X^{(t)}(-\tau) &=& \int_{-\tau}^0{d\sigma'\delta (\sigma'+\tau) X^{(t)}(\sigma')}
\end{eqnarray}
using Dirac's delta function $\delta(\cdot)$, we obtain
\begin{equation}
{\bf \bar{\Omega}}^{(t)}(\sigma) = \mathcal{N}_1(t)\delta(\sigma) + \mathcal{N}_2(t)\delta(\sigma+\tau),
\end{equation}
where $\mathcal{N}_j(t) \equiv \partial_{x_j}\mathcal{N}(x_1,x_2)$ $(j=1,2)$ is evaluated at $(x_1,x_2)=(X_0^{(t)}(0),X_0^{(t)}(-\tau))$.
The linearized dynamics for the deviation $Y(t)$ can then be written as
\begin{eqnarray}
L^{(t)}\left(Y^{(t)}\right)(\sigma)=
\left\{ \begin{array}{lr}
{\displaystyle \frac{d}{d\sigma}Y^{(t)}(\sigma)} & (-\tau\leq\sigma<0),\\
\\
{\displaystyle \mathcal{N}_1(t)Y^{(t)}(0) + \mathcal{N}_2(t)Y^{(t)}(-\tau)} & (\sigma=0).
\end{array}\right.
\end{eqnarray}

Let us introduce a time-periodic linear operator $\hat{L}$ of period $T$, which acts on a complexified Banach space $(C_0)_{\mathbb{C}}$~\cite[Sec. III.7]{Diekmann} as 
\begin{equation}
\left( \hat{L}Y^{(t)} \right)(\sigma) \equiv -\frac{d}{dt}Y^{(t)}(\sigma)+L^{(t)}\left(Y^{(t)}\right)(\sigma) \quad (-\tau\leq\sigma\leq0), \label{eq. hatL}
\end{equation}
and rewrite Eq.~(\ref{eq. var}) as $( \hat{L} Y^{(t)} )(\sigma) = 0$.
Because $Y^{(t)}$ obeys a periodically driven linear system, by the Floquet theorem for linear DDEs~\cite{Stokes, Hale, Simmendinger}, the spectrum of $\hat{L}$ is at most countable and
\begin{equation}
\left(\hat{L}q_{i}^{(t)}\right)(\sigma)=\lambda_{i}q_{i}^{(t)}(\sigma)\label{3}
\quad (-\tau\leq\sigma\leq0)
\end{equation}
is satisfied, where $\lambda_{i} \in {\mathbb C}$ is the $i$-th Floquet eigenvalue and $q_{i}^{(t)}\in (C_0)_{\mathbb C}$ is the corresponding $T$-periodic Floquet eigenfunction ($i=0, 1, 2, ...$).
Here, the largest eigenvalue, which is $0$ and simple by the Floquet theorem, is denoted as $\lambda_0=0$ and the other eigenvalues are arranged in descending order of the real part.

We also introduce adjoint eigenfunctions with respect to a bilinear form appropriate for DDEs~\cite{bilinear}. 
Following Refs.~\cite{Stokes, Hale, Simmendinger}, we define a bilinear form of two functions, $A \in (C_{0})_{\mathbb{C}}$ and $B\in (C_{0})_{\mathbb{C}}^{*}$, as
\begin{eqnarray}
\langle{B^{(t)}}, A^{(t)};t\rangle\equiv \left[B^{(t)}(0), A^{(t)}(0)\right]-\int_{-\tau}^{0}d\sigma\int_{0}^{\sigma}d\xi\left[B^{(t)}(\xi-\sigma),\ {\bf \bar{\Omega}}^{(t+\xi-\sigma)}(\sigma)A^{(t)}(\xi)\right]\label{bilinear}.
\end{eqnarray}
Here, $(C_{0})_{\mathbb{C}}^{*}=C([0,\tau]\to \mathbb{C}^{N*} )$ is the dual space of $(C_{0})_{\mathbb{C}}$ with respect to the bilinear form, consisting of (row) vector-valued functions that map the interval $[0,\tau]$ to $\mathbb{C}^{N*}$,
and $[\cdot,\cdot]$ denotes the Hermitian scalar product 
of $V\in  \mathbb{C}^{N*}$ and $U\in  \mathbb{C}^{N}$
defined as 
$[V,U] = \sum_{k=1}^N V_k \overline{U_k}$ where $V_k$ and $U_k$ are vector components of $V$ and $U$, respectively.
An adjoint operator $\hat{L}^{*}$ of $\hat{L}$ with respect to this bilinear form can then be derived as
\begin{eqnarray}
\left( \hat{L}^{*}Y^{(t)*} \right)(s)=\frac{d}{dt}Y^{(t)*}(s)+L^{(t)*}\left(Y^{(t)*}\right)(s)
\quad
(0\leq s\leq\tau) \label{3*},
\end{eqnarray}
where
\begin{eqnarray}
L^{(t)*}\left(Y^{(t)*}\right)(s)=
\left\{ \begin{array}{lc}
{\displaystyle -\frac{d}{ds}Y^{(t)*}(s)} & (0<s\leq\tau),\\
\\
{\displaystyle \int_{0}^{\tau}ds'Y^{(t)*}(s'){\bf \bar{\Omega}}^{(t+s')}(-s')} & (s=0).
\end{array}\right.\label{3*-1}
\end{eqnarray}
Here, $Y^{*}(t)\ \in \mathbb{C}^{N*}$ is a row vector of $N$ complex components
and $Y^{(t)*}(s)\equiv Y^{*}(t+s)$ $(0\leq s\leq\tau)\in (C_{0})_{\mathbb{C}}^{*}$
is its history-function representation.

The adjoint eigenfunction $q_{i}^{(t)*} \in (C_0)^*_{\mathbb C}$ of $q_{i}^{(t)}$, which is also $T$-periodic, satisfies
\begin{equation}
\left(\hat{L}^{*}q_{i}^{(t)*}\right)(s)=\bar{\lambda}_{i}q_{i}^{(t)*}(s) 
\quad
(0\leq s\leq\tau), \label{3*-2}
\end{equation}
where $\bar{\lambda}_{i}$ is the complex conjugate of ${\lambda}_{i}$.
If $\lambda_i \neq \lambda_j$, $q_{i}^{(t)}$ is orthogonal to $q_{j}^{(t)*}$ with respect to the bilinear form Eq.~(\ref{bilinear}), and 
hence they can be normalized to satisfy the biorthogonal relation 
$\langle{q_i^{(t)*}}, q_j^{(t)};t\rangle = \delta_{i, j}$.
The zero eigenfunction of the linear operator $\hat{L}$ can be chosen as $q_0^{(t)}(\sigma)=dX_{0}/dt|_{t+\sigma}$ ($-\tau\leq\sigma\leq0$), which can be confirmed by differentiating Eq.~(\ref{2}) with respect to $t$ at $X^{(t)} = X_{0}^{(t)}$ on the periodic orbit~\cite{Kotani}.
Note that this definition specifies the normalization of $q_0^{(t)}$.
For the other eigenfunctions $q_{i}^{(t)}$ ($i=1, 2, ...$), we normalize them such that $\max_{0 \leq t \leq T} \left( q_{i}^{(t)}(0)\right) =1$.
We use this convention for the normalization throughout this study.
We note that the zero eigenfunction of the linear operator $\hat{L}$ corresponds to the tangential component along the limit cycle, namely, the phase direction. Moreover,  the zero eigenfunction $q_{0}^{(t)*}$ of the adjoint operator
$\hat{L}^{*}$ gives the {\it phase sensitivity function} of the limit cycle, which characterizes linear response property of
the oscillator phase to weak perturbations~\cite{Kotani,Pyragas}. 
Similarly, the other eigenfunctions $q_{i}^{(t)*} $ ($i=1, 2, ...$) characterize linear response properties of the amplitudes and called {\it isostable response curves} for the case of ODEs~\cite{ErmentroutPTRSA}.

The adjoint eigenfunctions can numerically be obtained by an extension of the adjoint method for DDEs, which was previously used to calculate the adjoint zero eigenfunction of $\hat{L}$~\cite{Kotani,Pyragas}.
That is, we numerically integrate the linearized and its adjoint equations while subtracting unnecessary functional components by using the biorthogonality between the eigenfunctions and adjoint eigenfunctions.
The main difference from the adjoint method for $q_{0}^{(t)*}$ developed in the previous studies is that we calculate the adjoint eigenfunctions also for $\lambda_i \: (i\geq 1)$.
Therefore, during numerical integration, we need to remove unnecessary functional components in the directions of the lower-order eigenfunctions from $0$-th to $(i-1)$-th, which grow faster than the $i$-th component in order to calculate the $i$-th eigenfunction precisely. For $i \geq 1$, we also need to renormalize the solutions of the equations by a factor $e^{\lambda_i t}$ determined by the Floquet exponent in order to obtain the correct eigenfunctions.

To numerically calculate the $i$-th eigenfunction $q_{i}^{(t)}$, we integrate the linearized equation
\begin{align}
\frac{d}{dt}Y^{(t)}(\sigma)=L^{(t)}\left(Y^{(t)}\right)(\sigma)
\quad (-\tau\leq\sigma\leq0)
\label{eq:linearizedeq}
\end{align}
forward in time. During the calculation, we subtract the $0$-th to $(i-1)$-th eigencomponents from the numerical $Y^{(t)}$, which are unnecessary but arises due to numerical errors. The Floquet eigenvalue $\lambda_i$ is numerically evaluated from the exponential decay rate of $Y^{(t)}$. Then the eigenfunction $q_{i}^{\left(t\right)}\left(\sigma\right)$ is obtained by compensating the exponential decay of $ Y^{(t)}(\sigma)$ as $q_{i}^{\left(t\right)}\left(\sigma\right)=e^{-\lambda_{i}t} \ Y^{(t)}(\sigma)$ ($-\tau\leq\sigma\leq0$). See Sec. III.C and Ref. \cite{footnotesec3c} for further details.
In a similar way, the $i$-th adjoint eigenfunction $q_{i}^{(t)^*}$ is calculated by numerically integrating the adjoint linear equation
\begin{align}
\frac{d}{dt}Y^{(t)*}(s)=-L^{(t)*}\left(Y^{(t)*}\right)(s)
\quad
(0\leq s\leq\tau)
\label{eq:adjoint}
\end{align}
backward in time while subtracting unnecessary eigencomponents and then compensating the numerical result by a factor $e^{- \lambda_i t}$.
We call this procedure the extended adjoint method in this study.

\subsection{Nonlinear phase-amplitude equations \label{sec. pa}}

Our aim is to derive a set of low-dimensional dynamical equations from the original DDE by projecting the system state onto a moving coordinate frame spanned by
a small number of Floquet eigenfunctions.
That is, we decompose the deviation of the system state $X^{(t)}$ from that on the limit cycle $X_0^{(t)}$ by using the eigenfunctions associated with the leading $M$ eigenvalues other than $0$, which are assumed to be real and simple for the sake of simplicity~\cite{lambda_complex}, as
\begin{align}
X^{(t)}(\sigma) \simeq X_0^{(\phi)}(\sigma) + \sum_{i=1}^{M} \rho_i(t) q_i^{(\phi)}(\sigma),
\quad
(-\tau\leq\sigma\leq0),
\end{align}
%
%
where $X_0^{(\phi)}$ is a system state on the limit cycle parametrized by the phase $\phi \in\left[0,T\right)$, $q_i^{(\phi)}$ ($i=1, ..., M$) is the Floquet eigenfunction associated with $\lambda_i$ and denoted as a function of $\phi$ rather than $t$,
and $\{ \rho_i(t) \}$  are real expansion coefficients representing amplitudes of the Floquet eigenmodes. 
The symbol $\simeq$ indicates that we approximate $X^{(t)}(\sigma)$ by its projection on the space spanned by the $M$ eigenfunctions $\{ q_1^{(\phi)}, ... ,q_M^{(\phi)}\} $. 
We here use the term ``amplitude'' in a generalized sense, allowing it to take both positive and negative values; it is the component of the system state along the Floquet eigenfunction corresponding to the direction transversal to the limit cycle and represents the deviation of the system state from the limit cycle.
%
Here, the phase value $\phi$ for a given state $X^{(t)}$ is determined in such a way that the state difference $X^{(t)} - X_0^{(\phi)}$ does not have a tangential functional component $q_0^{(\phi)}$ along the limit cycle. Thus, we assume the following orthogonality condition:
\begin{equation}
\left\langle q_{0}^{\left(\phi\right)*}, X^{(t)}-X_{0}^{\left({\phi}\right)};{\phi}\right\rangle = 0,
\label{orthogonality}
\end{equation}
namely, the difference $X^{(t)} - X_0^{(\phi)}$ is on the hyperplane that is tangent to the isochron on the limit cycle at $X_0^{(\phi)}$.
Note that the phase defined in this way is different from the asymptotic phase.
%

Because we use a linear coordinate frame spanned by the Floquet eigenfunctions $\{ q_i ^{(\phi)} \}$ ($i=1, ..., M$), 
nonlinear interactions between different eigenmodes generally arise.
Specifically, when the eigenvalue $\lambda_{1}$ with the largest non-zero part is close to $0$, the perturbed system state does not go back to the limit cycle quickly, and hence nonlinear interactions between the phase eigenmode and the slowest-decaying amplitude eigenmode should be taken into account for better description of the system.

For ordinary differential equations, such coupled nonlinear phase-amplitude equations have been derived by transforming the original equations around the limit cycle in several contexts~\cite{Wedgwood,Morita}.
Such transformation methods have also been developed for DDEs in Refs.~\cite{Stokes2,Hale2}, though the treatments of oscillatory dynamics in these studies are rather abstract.
Quantitative analysis of synchronization dynamics of DDEs using the coordinate transform proposed therein have not been very fruitful despite their potential advantages, mainly due to the lack of practical methods for numerically evaluating the Floquet eigenfunctions.

We hereafter restrict ourselves to the case in which $\lambda_1$ takes a negative real value near zero and $\mbox{Re}\{\lambda_2\}\ll \lambda_1$ for simplicity.
To derive the phase and amplitude equations, we retain only the slowest two modes associated with $\lambda_0$ and $\lambda_1$ and approximate $X^{(t)}({\sigma})$ as 
\begin{align}
X^{(t)}({\sigma}) \simeq X_{0}^{(\phi)}({\sigma}) + R(t) q_{1}^{(\phi)}(\sigma),
\end{align}
where 
$R(t) = \rho_1(t)$ is the amplitude of the eigenmode corresponding to $\lambda_1$.
The symbol $\simeq$ here indicates that we further approximate $X^{(t)}(\sigma)$ by its projection on a one-dimensional space spanned by $q_1^{(\phi)}$.
We substitute this expression into Eq.~(\ref{2}) and then project both sides of Eq.~(\ref{2})
onto the eigenfunctions $q_{0}^{(\phi)}$ and  $q_{1}^{(\phi)}$, respectively, by using biorthogonality of the eigenfunctions and derive the equations for the phase $\phi$ and the amplitude $R$.

As explained in Appendix~\ref{app. pae}, the phase equation can be derived as 
\begin{eqnarray}
\frac{d\phi}{dt}&=&1+\frac{q_{0}^{(\phi)*}\left(0\right)\cdot\left(F_{\mathrm{nl}}\left(\phi,R\right)+G\left(\phi,R,t\right)\right)}{1+R\left\langle q_{0}^{(\phi)*},L^{(\phi)}\left(q_{1}^{(\phi)}\right);\phi\right\rangle },
\label{eq:13before_combined}
\end{eqnarray}
\color{black}
or, by rewriting the right-hand side,
\begin{align}
\frac{d\phi}{dt}
=&
1
+ q_{0}^{(\phi)*}\left(0\right)\cdot\left(F_{\mathrm{nl}}\left(\phi,R\right)+G\left(\phi,R,t\right)\right)
\cr
&-\frac{R\left\langle q_{0}^{(\phi)*},L^{(\phi)}\left(q_{1}^{(\phi)}\right);\phi\right\rangle}{1+R\left\langle q_{0}^{(\phi)*},L^{(\phi)}\left(q_{1}^{(\phi)}\right);\phi\right\rangle }
q_{0}^{(\phi)*}\left(0\right)\cdot\left(F_{\mathrm{nl}}\left(\phi,R\right)+G\left(\phi,R,t\right)\right),
\label{eq:13before}
\end{align}
and the amplitude equation can similarly be derived as
\begin{align}
\frac{dR}{dt}=&\lambda_{1}R
+q_{1}^{(\phi)*}\left(0\right)\cdot\left(F_{\mathrm{nl}}\left(\phi,R\right)+G\left(\phi,R,t\right)\right)
\cr
&-\frac{ R\left[\left\langle q_{1}^{(\phi)*}, L^{(\phi)}\left(q_{1}^{(\phi)}\right);\phi\right\rangle - \lambda_{1} \right] }{1+R\left\langle q_{0}^{(\phi)*},L^{(\phi)}\left(q_{1}^{(\phi)}\right);\phi\right\rangle } q_{0}^{(\phi)*}\left(0\right)\cdot\left(F_{\mathrm{nl}}\left(\phi,R\right)+G\left(\phi,R,t\right)\right),
\label{eq:16before}
\end{align}
where the nonlinear functional $\mathcal{N}$ in Eq.~(\ref{eq:taylorexpandN}) is approximated by an ordinary function of $\phi$ and $R$,
\begin{eqnarray}
F_{\mathrm{nl}}\left(\phi,R\right)
&\equiv&
F_{\mathrm{nl}} \left( R q_{1}^{(\phi)} (\cdot) \right)
=
\mathcal{N}\left(X_{0}^{(\phi)} (\cdot) + R q_{1}^{(\phi)} (\cdot)\right)-\mathcal{N}\left(X_{0}^{(\phi)} (\cdot) \right)
-L^{(\phi)}\left( R q_1^{(\phi)} \right)(0),
\label{eq:def_fnl}
\end{eqnarray}
and the external perturbation is also approximated as
\begin{align}
G(\phi, R, t) \equiv G\left(X_0^{(\phi)} (\cdot) + R q_1^{(\phi)} (\cdot) , t\right).
\label{eq:G}
\end{align}
In Eq.~(\ref{eq:13before}) and Eq.~(\ref{eq:16before}), both the second and third terms on the right-hand side depend on $F_{\mathrm{nl}}$ and $G$.
Note that $F_{\mathrm{nl}}(\phi, R)$ includes only terms of $O(R^2)$ or higher, because the constant terms and linear terms in $R$ have already been subtracted in Eq.~(\ref{eq:def_fnl}). 

Thus, by projecting the DDE onto the first two eigenfunctions, a set of two-dimensional coupled ordinary differential equations for the phase $\phi$ and amplitude $R$ is obtained.
In order to consider the higher-order effects of the amplitude deviations, we have not expanded the third-order terms in Eq.~(\ref{eq:13before}) and Eq.~(\ref{eq:16before}) in a series of $R$ and hence the dynamics of $\phi$ and $R$ are nonlinearly coupled at the second and higher orders in $R$.
This nonlinearity can be a source of intriguing oscillatory dynamics~\cite{Canavier+,Rubin+,Castejon+, Rabinovitch+}.
We also note that the lowest-order phase-amplitude equations (see Refs.~\cite{Castejon+, ErmentroutPTRSA, KuramotoPTRSA} for the case of ODEs) 
\begin{align}
\frac{d\phi}{dt}
=&
1
+ q_{0}^{(\phi)*}\left(0\right)\cdot G\left(\phi,R,t\right) 
\cr
\frac{dR}{dt}
=&\lambda_{1}R
+q_{1}^{(\phi)*}\left(0\right)\cdot G\left(\phi,R,t\right) 
\end{align}
are obtained at the lowest-order approximation in $R$, where $F_{\mathrm{nl}}\left(\phi,R\right)$ is $O(R^2)$ and does not appear at the lowest order.

Finally, before proceeding, we note that there are also other formulations of phase or phase-amplitude reduced equations for analyzing higher-order effects of perturbations on limit cycles described by ODEs, such as non-pairwise phase  interactions~\cite{KuramotoPTRSA}, higher-order phase reduction~\cite{Pazo}, nonlinear phase coupling function~\cite{RosenblumPTRSA}, and higher-order approximations of coupling functions~\cite{ErmentroutPTRSA}, which can capture more detailed aspects of synchronization than the lowest-order phase equation.

\subsection{Averaged phase-amplitude equations}

When the perturbation applied to the oscillator is a periodic external force whose frequency is close to the natural frequency of the oscillator, we may further derive simpler, approximate phase-amplitude equations by averaging out the fast oscillatory component as follows.

We assume that the perturbation $G$ is purely external (i.e. independent of the system state and periodic in $t$ with period $T' = T/r$ (frequency $r$), i.e.,  %
\begin{align}
G(t + T/r) = G(t).
\end{align}
We also assume that the detuning between the natural frequency of the oscillator and the periodic force is small and denote it as $\Delta\omega = 1 - r$. 

We introduce a slow phase variable $\psi \equiv \phi - r t$. The equations for $\psi$ and $R$ can be written as
\begin{eqnarray}
\frac{d\psi}{dt} &=& \Delta \omega + \frac{q_{0}^{(\psi+rt)*}\left(0\right)\cdot\left(F_{\mathrm{nl}}\left(\psi+rt, R\right)+G\left(t\right)\right)}{1+R\left\langle q_{0}^{(\psi+rt)*},L^{(\psi+rt)}\left(q_{1}^{(\psi+rt)}\right);\psi+rt\right\rangle },
\label{eq:psinonavg}
\end{eqnarray}
and
\begin{align}
\frac{dR}{dt}=&\lambda_{1}R
+q_{1}^{(\psi+rt)*}\left(0\right)\cdot\left(F_{\mathrm{nl}}\left(\psi+rt,R\right)+G\left(\psi+rt,R,t\right)\right)
\cr
&-\frac{ R\left[ \left\langle q_{1}^{(\psi+rt)*}, L^{(\psi+rt)}\left(q_{1}^{(\psi+rt)}\right);\psi+rt\right\rangle -\lambda_{1} \right] }{1+R\left\langle q_{0}^{(\psi+rt)*},L^{(\psi+rt)}\left(q_{1}^{(\psi+rt)}\right);\psi+rt\right\rangle } q_{0}^{(\psi+rt)*}\left(0\right) 
\cr
&\cdot\left(F_{\mathrm{nl}}\left(\psi+rt,R\right)+G\left(\psi+rt,R,t\right)\right).
\label{eq:Rnonavg}
\end{align}
We also expand the nonlinear term $F_{\mathrm{nl}}$ in Taylor series of $R$ up to $R^N$ as
\begin{align}
F_{\mathrm{nl}}(\psi + rt, R) &= \sum_{\ell=2}^{N} R^\ell F_{\mathrm{nl},\ell}(\psi + rt) + O(R^{N+1}),
\end{align}
where $\{ F_{\mathrm{nl},\ell} \}$ ($\ell=2, 3, ...$) are expansion coefficients. 
Note that the series for $F_{\mathrm{nl}}$ starts from $O(R^2)$.

Considering that $\psi$ evolves only slowly while $r t$ rapidly increases, we approximate the terms with $\psi+rt$ in Eqs.~(\ref{eq:psinonavg}) and (\ref{eq:Rnonavg}) by their one-period average, for example, as 
\begin{align}
q_0^{(\psi+rt)*}(0) \cdot F_{\mathrm{nl},2}(\psi + rt)
\approx 
\frac{1}{T'} \int_0^{T'} q_0^{(\psi+r s)*}(0) \cdot F_{\mathrm{nl},2}(\psi + rs) ds
= \frac{1}{T} \int_0^{T} q_0^{(\theta)*}(0) \cdot F_{\mathrm{nl},2}(\theta) d\theta
= a_1
\end{align}
and
\begin{align}
q_0^{(\psi+rt)*}(0) \cdot G(t)
\approx 
\frac{1}{T'} \int_0^{T'} q_0^{(\psi+r s)*}(0) \cdot G(s) ds
=
\frac{1}{T} \int_0^{T} q_0^{(\theta)*}(0) \cdot G\left( \frac{\theta - \psi}{r} \right) d\theta
=
g_0(\psi),
\end{align}
where $\psi$ is kept constant during the integration.
Expanding $F_{\mathrm{nl}}(\psi+rt, R)$ up to $O(R^3)$ and averaging the coefficients,
 we obtain approximate equations for $\psi$ and $R$ as

\begin{equation}
\frac{d\psi}{dt}=\Delta\omega+\frac{1}{1+Ra_{0}}\left(a_{2}R^{2}+a_{3}R^{3}+g_0(\psi) \right),
\label{eq:psi_expand}
\end{equation}
and
\begin{align}
\frac{dR}{dt} = \lambda_1 R
+b_{2}R^{2}+b_{3}R^{3}- \frac{R(b_0 - \lambda_1)}{1+Ra_{0}}\left(a_{2} R^{2} + a_{3}R^{3} + g_0(\psi) \right)
+g_1(\psi),
\label{eq:R_expand}
\end{align}
where the equations for the individual coefficients are given in Appendix~\ref{app:coefficients}.
We check the validity of the above averaging approximation numerically in the next section.

\subsection{Evaluation of the phase and amplitude}

Numerically, the values of the phase $\phi$ and amplitude $R$ can be evaluated from the system state $X^{(t)}$ by the following two-step procedure.
First, we evaluate the phase of the state $X^{(t)}$ by
choosing the phase value $\phi$ so that it satisfies the orthogonality condition Eq.~(\ref{orthogonality}).
Numerically, we find the value $\hat{\phi}$ that minimizes the mean squared error,
\begin{align}
\left| \left\langle q_{0}^{\left(\hat{\phi}\right)*}, X^{(t)}-X_{0}^{\left(\hat{\phi}\right)};\hat{\phi}\right\rangle \right|^2.
\end{align}
There exists a neighborhood of the periodic orbit where the phase and amplitude components defined by using the Floquet eigenfunctions are uniquely determined~\cite[Lemma1]{Hale2}.
However, in general, there can exist multiple values of $\hat{\phi}$ that satisfy Eq.~(\ref{orthogonality}) in the range $0\leq\hat{\phi}<T$.
To choose the appropriate value from them, for each candidate of $\hat{\phi}$, we evaluate the corresponding $q_{1}$ component as
\begin{equation}
\hat{R} = \left\langle q_{1}^{\left(\hat{\phi}\right)*}, X^{(t)}-X_{0}^{\left(\hat{\phi}\right)};\hat{\phi}\right\rangle
\label{orthogonality_i}
\end{equation}
and adopt the value of $\hat{\phi}$ that has the smallest $\left|\hat{R}\right|$ as the estimate of $\phi$,
and the smallest $\hat{R}$ as the estimate of $R$.

\subsection{Approximate evaluation of the asymptotic phase}

The phase $\phi$ defined by the Floquet eigenfunction, which we use in the present study for the phase-amplitude description, is different from the asymptotic phase $\Phi$; the isosurface of $\Phi$ is generally curved and tangent to the isophase plane of $\phi$ at each point on the limit cycle.
Since the asymptotic phase $\Phi$ provides useful information on the nonlinear dynamical properties of the oscillator, it is convenient if we can approximate $\Phi$ using $\phi$ and $R$.
In this subsection, we propose a method to approximately evaluate the asymptotic phase of an unperturbed oscillator from $\phi$ and $R$ defined by the Floquet eigenfunctions, which is valid when $R$ is sufficiently small.

When the perturbation is absent ($G=0$), Eq.~(\ref{eq:13before_combined}) for $\phi$ can be written as
\begin{align}
\frac{d\phi}{dt} = 1+ d(\phi, R)
\label{eq:R1}
\end{align}
where 
\begin{align}
d(\phi, R) = \frac{q_{0}^{(\phi)*}\left(0\right) \cdot F_{\mathrm{nl}}\left(\phi,R\right) }{1+R\left\langle q_{0}^{(\phi)*},L^{(\phi)}\left(q_{1}^{(\phi)}\right);\phi\right\rangle }.
\end{align}
The asymptotic phase $\Phi$ of the system state $X^{(t_0)}$ at time $t_0$ with phase $\phi_0$ and amplitude $R_0$ can approximately be obtained by integrating $d(\phi(s), R(s))$ until the system state goes back sufficiently close to the limit cycle as 
%
\begin{align}
\Phi (X^{(t_0)}) = \phi_0 + \int_{t_0}^\infty d(\phi(s), R(s)) ds.
\end{align}

When $R$ is sufficiently small, we may ignore the higher-order terms in $R$ in the equations for $\phi$ and $R$ and assume that $\phi$ increases constantly with frequency $1$ and $R$ decays exponentially with rate $\lambda_1$ as 
\begin{eqnarray}
\phi=\phi_0 + t - t_0,
\quad
R \left(t\right)=R_0 \exp\left(\lambda_1 (t-t_0) \right),
\label{eq:R2}
\end{eqnarray}
at the lowest-order approximation. The asymptotic phase $\Phi$ of the system state $X^{(t_0)}$ can then be approximately evaluated as 
\begin{eqnarray}
\hat{\Phi} = \phi_0+\int_{t_0}^{\infty} d\left( \phi_0 + s - t_0, R_0 \exp\left(\lambda_1 (s-t_0) \right) \right) ds.
\label{eq:R3}
\end{eqnarray}
In Sec.~III~E and Sec.~IV, we use the above method to estimate the asymptotic phase $\Phi$ of the oscillator and compare it with direct numerical results.

\section{Analytically tractable model}

To demonstrate the validity of the proposed framework, we first consider a limit-cycle oscillator described by a scalar DDE with a cubic nonlinearity, for which approximate expressions of the Floquet eigenfunctions and their adjoints can be analytically derived, and analyze the effect of a periodic force on the dynamics.

\subsection{Model}

The model is represented as
\begin{equation}
\frac{dx(t)}{dt}=-x\left( t-\frac{\pi}{2} \right) + \epsilon x(t) \left[ 1-x(t)^{2}-x \left(t-\frac{\pi}{2}\right)^{2} \right] + G(t),\label{X}
\end{equation}
where $x(t)\in \mathbb{R}$, $\epsilon=0.05$ is a small constant, and the external periodic force is described by
\begin{align}
G(t)=G_{0}\sin\left(\frac{2\pi}{T} r t \right),
\end{align}
where $G_0$ is the intensity of the periodic force and $r$ is the ratio of the natural frequency $2\pi/T$ of the limit cycle to that of the external force.
It is assumed that $r$ is sufficiently close to $1$.

When $G = 0$, this DDE has a limit cycle of period $T=2\pi$ given by $x_{0}(t)=\sin t$, or
\begin{align}
x_{0}^{(t)}(\sigma) =\sin ( t + \sigma )
\quad
(-\tau\leq\sigma\leq0)
\end{align}
in the history-function representation, and its rate of attraction to the limit cycle is determined by $\epsilon$.
When $\epsilon$ is small, the relaxation time of the system state to the limit cycle is considerably large as compared to the oscillation period as shown in Figs.~\ref{figs}(a) and (b).

We denote the small deviation of the system state from the limit cycle as $y^{(t)}(\sigma)=x^{(t)}(\sigma)-x_{0}^{(t)}(\sigma)$ $(-\tau \leq \sigma \leq 0)$.
The linear operator $\hat{L}$ of this system is given by Eq.~(\ref{2-1}) with
\begin{equation}
{\bf \bar{\Omega}}^{(t)}(\sigma)=\delta(\sigma) \left[ \epsilon(1-3x_{0}(t)^{2} - x_{0}\left(t-\frac{\pi}{2}\right)^{2} \right]
- \delta\left(\sigma+\frac{\pi}{2}\right) \left[ 1+2\epsilon x_{0}(t) x_{0}\left(t-\frac{\pi}{2}\right) \right],
\label{Omega}
\end{equation}
where $\delta$ is Dirac's delta function.
By retaining the first two leading eigenvalues, the nonlinear phase-amplitude equations can be derived as Eqs.~(\ref{eq:13before}) and (\ref{eq:16before}).

\subsection{{Approximate analytical expressions of the eigenvalues and eigenfunctions}
\label{app. exadj}}

We first derive approximate Floquet eigenvalues, eigenfunctions, and adjoint eigenfunctions of the model Eq.~(\ref{X}) without the external force ($G=0$) analytically.
In what follows, we consider the case in which the relaxation of the system state to the limit cycle is slow and assume that $\lambda_{1}$ is small and $O(\epsilon)$.
First, the zero eigenfunction of $\hat{L}$ is given exactly as 
\begin{align}
q_{0}^{(t)}(\sigma)=\cos(t+\sigma)
\quad
(-\tau\leq\sigma\leq0)
\end{align}
and the adjoint eigenfunction is
\begin{align}
q_{0}^{(t)*}(s)=\frac{8}{\epsilon\pi+4}\cos(t+s)
\quad
(0\leq s\leq\tau).
\end{align}
To find the exponent $\lambda_1$ with the second largest real part, we introduce an ansatz
\begin{align}
q_{1}^{(t)}(\sigma)=Ce^{\lambda_{1}\sigma}(\sin(t+\sigma)+l\cos(t+\sigma))
\quad
(-\tau\leq\sigma\leq0)
\end{align}
where $l$ is a constant and plug this into Eqs.~(\ref{eq. var}) and (\ref{2-1}).
We then obtain the approximate eigenvalue and the associated eigenfunction up to $O(\epsilon)$ as
\begin{align}
\lambda_{1}=-\frac{8\epsilon}{\pi^{2}+4}
\end{align}
and
\begin{align}
q_{1}^{(t)}(\sigma)=\frac{2}{\sqrt{4+\pi^{2}}}e^{-\frac{8\epsilon\sigma}{\pi^{2}+4}}\left( \sin(t+\sigma)-\frac{\pi}{2} \cos(t+\sigma) \right)
\quad
(-\tau\leq\sigma\leq0),
\end{align}
respectively.
Similarly, for the corresponding adjoint eigenfunction, we approximately obtain 
\begin{align}
q_{1}^{(t)*}(s)=C_{1}^{*} e^{\frac{8\epsilon s}{\pi^{2}+4}} \left( \sin(t+s)+\frac{\pi}{2} \cos(t+s) \right)
\quad
(0\leq s\leq\tau),
\end{align}
where the constant $C_{1}^{*}$ is determined from the normalization
condition $\langle q_{1}^{(t)*},q_{1}^{(t)};t\rangle=1$.

\subsection{Numerical evaluation of the eigenvalues and eigenfunctions}

To confirm the validity of the approximate analytical results for the Floquet eigenvalues, eigenfunctions, and adjoint eigenfunctions obtained in the previous subsection, we numerically evaluate these quantities by the extended adjoint method and compare with the approximate analytical results.

First, as in the conventional adjoint method for DDEs~\cite{Kotani,Pyragas}, we compute $q_{0}^{(t)}(\sigma)$ $(-\tau\leq\sigma\leq0)$, which is simply $dX_0/dt|_{t+\sigma}$, and then $q_{0}^{(t)*}(\sigma)$ $(-\tau\leq\sigma\leq0)$ by backwardly integrating the adjoint linear equation. The adjoint eigenfunction $q_{0}^{(t)*}$ is normalized such that $\langle q_{0}^{(t)*},q_{0}^{(t)};t\rangle=1$.
Next, we obtain the eigenfunction $q_{1}^{(t)}$ with the largest negative eigenvalue ($\lambda_{1}<0$, $\lambda_{1}>\lambda_{i}$ for $i=2,\cdots ,M$)~\cite{lambda_complex}.
As an initial function, we take an arbitrary function $Y_{\mathrm{ini}}^{(t=0)}$ at $t=0$~\cite{comment0}, subtract the $q_{0}^{(t=0)}$ component from this initial function as $Y^{(t=0)}(\sigma) = Y_{\mathrm{ini}}^{(t=0)}(\sigma)-\langle q_{0}^{(0)}{}^{*},Y_{\mathrm{ini}}^{(0)};0\rangle q_{0}^{(0)}(\sigma)$ ($-\tau\leq\sigma\leq0$), where the second term represents the projection of $Y_{\mathrm{ini}}^{(t=0)}$ onto $q_0^{(0)}$, and numerically integrate the linear equation~(\ref{eq:linearizedeq}) for $Y^{(t)}$ from this initial condition to $t=T$ as explained before.

Similarly, in order to compute the eigenfunction $q_{1}^{(t)*}$, we initialize $Y^{(t=0)*}(s)$  ($0\leq s\leq \tau$) appropriately and numerically integrate Eq.~(\ref{eq:adjoint}) backward in time, subtracting the $q_{0}^{(t)*}$ component at every period, and compensate the exponential decay in the numerical solution.
The adjoint eigenfunction $q_{1}^{(t)*}$ is normalized so that $\langle q_{1}^{(t)*},q_{1}^{(t)};t\rangle=1$.

\begin{figure}[th]
\begin{centering}
\includegraphics{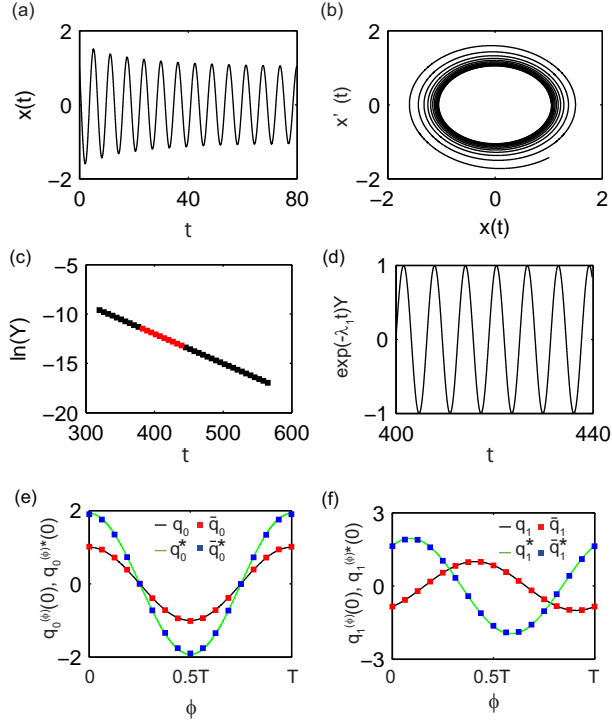}
\par\end{centering}
\caption{
 (a) Time course of a scalar DDE with a cubic nonlinearity, Eq.~(\ref{X}), showing a slow convergence of the system state
to the limit cycle.
(b) Time course of the oscillator state projected
onto the $(x,dx/dt)$-plane.
[(c), (d)] Extended adjoint method for calculating $q_1^{(t)}$;
(c) Peak heights of the time course of $Y^{(t=nT)}(0)$ measured at each period vs. $t$.
The red squares are the data from which the Floquet exponent $\lambda_1$ is evaluated.
(d) Time evolution of $\exp\left(-\lambda_{1}t\right) Y^{(t)}(0)$ after compensating the exponential decay.
(e) Eigenfunctions and adjoint eigenfunctions associated with $\lambda_{0} = 0$ plotted as functions of $\phi$.
The functions $q_{0}^{(\phi)}(0)$ and $q_{0}^{(\phi)*}(0)$ are analytically derived, while $\bar{q}_{0}^{(\phi)}(0)$ and $\bar{q}_{0}^{(\phi)*}(0)$ are numerically  obtained by the extended adjoint method.
(f) Eigenfunctions $q_{1}^{(\phi)}(0)$ and adjoint eigenfunctions $q_{1}^{(\phi)^*}(0)$ associated with $\lambda_{1}$. }
\label{figs}
\end{figure}

Figure~\ref{figs}(c) shows the exponential decay of the peak heights of $Y^{(t=nT)}(0)$, from which we obtain the Floquet eigenvalue $\lambda_{1}$. Figure~\ref{figs}(d) shows  the time course of $e^{-\lambda_{1}t}\ Y^{(t)}(0)$ that is used for numerical computation of eigenfunction  $q_1^{(\phi)}$.
Figures~\ref{figs}(e) and (f) show the obtained pair of Floquet eigenfunctions, where $q_{0}^{(\phi)}(0)$ and $q_{0}^{(\phi)*}(0)$ are plotted with respect to $\phi$ in Fig.~\ref{figs}(e), and $q_{1}^{(\phi)}(0)$ and $q_{1}^{(\phi)*}(0)$ are plotted with respect to $\phi$ in Fig.~\ref{figs}(f).
We can confirm a good agreement between the numerical results and approximate analytical results for the eigenfunctions. The numerical value of the largest negative exponent $\lambda_{1}$ is approximately evaluated as $-0.030$, which is also close to the theoretical value $-8\epsilon/(\pi^{2}+4)=-0.029$.

\subsection{Phase-amplitude equations}

We now derive a set of nonlinear phase-amplitude equations from Eq.~(\ref{X}) with the periodic sinusoidal force.
The nonlinear term $F_{\mathrm{nl}}\left(\phi,R\right)$ in Eq.~(\ref{eq:def_fnl}) is explicitly given by
\begin{eqnarray}
F_{\mathrm{nl}}\left(\phi,R\right)
&=& \epsilon R q_{1}^{(\phi)}(0)  \left\{ -\left(x_{0}^{(\phi)}(0)+R q_{1}^{(\phi)}(0) \right)^2+\left(x_{0}^{(\phi)}(0)\right) ^2 
-\left( x_{0}^{(\phi)}\left(-\frac{\pi}{2}\right) + R q_{1}^{(\phi)} \left(-\frac{\pi}{2}\right)\right)^2
+\left(x_{0}^{(\phi)}\left(-\frac{\pi}{2}\right)\right)^2 \right\} \nonumber \\
&+& \epsilon x_{0}^{(\phi)}(0) \left(-\left(R q_{1}^{(\phi)}(0)\right) ^2- \left(R q_{1}^{(\phi)}\left(-\frac{\pi}{2}\right)\right) ^2\right)
\end{eqnarray}
and the reduced equations~(\ref{eq:13before}) and (\ref{eq:16before}) for $\phi$ and $R$ can be derived using this equation.

Expanding the nonlinear term $F_{\mathrm{nl}}$ and applying the averaging procedure, the approximate equations for the phase difference $\psi = \phi - rt$ and $R$ are given in the form of 
Eqs.~(\ref{eq:psi_expand}) and (\ref{eq:R_expand}) with
\begin{align}
g_0(\psi) = G_0 \frac{1}{T}\int_{0}^{T}q_{0}^{(\phi)*}\left(0\right)\cdot \sin \left( \frac{2\pi \left(\phi-\psi \right) }{T} \right) d\phi
=
G_0\left( g_{01} \sin \frac{2\pi \psi}{T} + g_{02} \cos \frac{2\pi \psi}{T} \right)
\label{eq:28-1}
\end{align}
and
\begin{align}
g_1(\psi) = \int_{0}^{T}q_{1}^{(\phi)*}\left(0\right)\cdot  \sin\left(2\pi \frac{\left(\phi-\psi \right)}{T} \right) d\phi
=
G_0 \left( g_{11} \sin \frac{2\pi \psi}{T} + g_{12} \cos \frac{2\pi \psi}{T} \right).
\end{align}

Using numerically evaluated eigenvalues and eigenfunctions, the coefficients in Eqs.~(\ref{eq:psi_expand}) and (\ref{eq:R_expand}) can be calculated as 
$\lambda_{1}=-0.029$, 
$a_{0}=1.8418$,
$a_{2}=0.0436$, 
$a_{3}=0.0415$; 
$b_{0}=1.5353$,
$b_{2}=-0.0053$, 
$b_{3}=0.0212$; 
and 
$g_{01}=-0.9622$, $g_{02}=0$, $g_{11} =-0.8239$, and $g_{12}=0.5245$.
From these coefficients, the equations for the phase difference $\psi$ and the amplitude $R$ are obtained as
\begin{align}
\frac{d\psi}{dt} &=\Delta\omega+\frac{1}{1+ 1.8418 R}\left(0.0436 R^{2}+0.0415 R^{3}  -0.9622  G_0\sin \frac{2\pi \psi}{T} \right),
\cr
\frac{dR}{dt} &= \lambda_1 R
-0.0053 R^{2}+ 0.0212 R^{3} -0.8239 G_0 \sin \frac{2\pi \psi}{T} + 0.5245 G_0 \cos \frac{2\pi \psi}{T}
\cr
&- \frac{R(1.5353-\lambda_1 )}{1+1.8418 R}\left(0.0436  R^{2} + 0.0415 R^{3}  -0.9622  G_0\sin \frac{2\pi \psi}{T} \right).
\label{eq:phaseamplitudeactual}
\end{align}
Thus, we have approximately reduced an infinite-dimensional dynamical system described by a DDE to a set of ODEs for the phase and amplitude.

\subsection{Approximate evaluation of the asymptotic phase}

In this subsection, we verify the validity of the approximate expression for the asymptotic phase derived in Sec.~II~F by evolving the present model from initial conditions far from the limit cycle.
From the reduced phase-amplitude equations and Eq.~(\ref{eq:R3}), the asymptotic phase $\Phi$ for the present model can be approximately evaluated from the phase $\phi$ and amplitude $R$ as
\begin{equation}
\hat{\Phi} = \phi+(0.7553 + 0.0448 \sin(2\phi+4.1499) + 0.0006 \sin(4\phi+2.6902)) R^{2}
\label{eq:34}
\end{equation}
up to $O(R^2)$. 
For a given system state $x^{(t)}$, the phase $\phi$ and amplitude $R$ can be evaluated as explained in Sec.~II~E, and the approximate asymptotic phase $\hat{\Phi}$ can then be obtained by Eq.~(\ref{eq:34}).
We also directly evaluate the asymptotic phase $\Phi$ for several initial conditions by numerically integrating the system and measuring the time necessary for the system state  to converge sufficiently close to the limit cycle for comparison.

\begin{figure}[H]
\begin{centering}
\includegraphics[width=8.6cm]{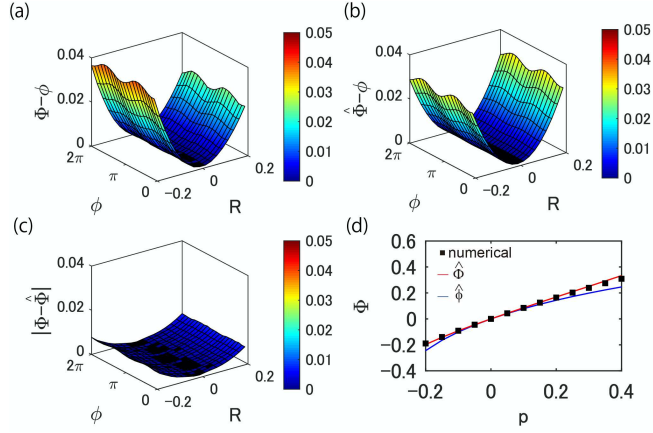}
\par\end{centering}
\caption{
Evaluation of the asymptotic phase $\Phi$ from $\phi$ and $R$.
(a) Difference $\phi - \Phi$ between $\phi$ and $\Phi$ plotted on the $(\phi, R)$-plane. The data are obtained by direct numerical integration
from initial system states given by $x^{(t=0)}(\sigma)=x_{0}^{(\phi)}(\sigma)+R q_{1}^{(\phi)}(\sigma)$.
(b) Difference $\phi - \hat{\Phi}$ between $\phi$ and $\hat{\Phi}$ estimated by using Eq.~(\ref{eq:34}). 
(c) Absolute difference $|\Phi - \hat{\Phi}|$ between the asymptotic phase $\Phi$ measured directly by numerical integration and $\hat{\Phi}$ estimated by using Eq.~(\ref{eq:34}). 
(d) Asymptotic phase of the initial system states given by $x^{(t=0)}(\sigma)=\sin\sigma+p\sin(\sigma/2)$,
which is not on the plane spanned by the first two Floquet eigenfunctions.
The black points indicate the numerical results, the blue line indicates
the phase $\phi$ evaluated using the linearized isochrons,
and the red line indicates the analytical estimation of asymptotic phase ${\Phi}$. 
}
\label{fig2}
\end{figure}

As the first example, we try to estimate $\Phi$ when the initial function is on the $\phi$-$R$ plane, that is, $x^{(t=0)}(\sigma)=x_{0}^{(\phi)}(\sigma)+R q_{1}^{(\phi)}(\sigma)$ ($-\tau\leq\sigma\leq0$).
Figure~\ref{fig2} (a) shows $\Phi - \phi$ for given initial values of $\phi$ and $R$ obtained by direct numerical integration of the DDE, and Fig.~\ref{fig2} (b) shows analytical results of $\hat{\Phi} - \phi$ obtained from Eq.~(\ref{eq:34}). Figure~\ref{fig2} (c) shows the absolute difference between $\Phi$ and its analytical estimation $\hat{\Phi}$. 
We can confirm a good agreement between the approximate analytical curve and direct numerical results for the whole range of $\phi$ when $|R|$ is not too large.

As the second example, we consider initial functions that are not on the $\phi$-$R$ plane. We set the initial functions as $x^{(t=0)}(\sigma)=\sin\sigma+p\sin(\sigma/2)$ $(-\tau \leq \sigma \leq0$)
with varying values of $p$~\cite{comment1}, and evaluated their asymptotic phase $\Phi$ by direct numerical integration of the DDE.
Figure~\ref{fig2} (d) shows the phase $\phi$, the asymptotic phase $\Phi$ estimated by Eq.~(\ref{eq:34}), and the asymptotic phase $\Phi$ obtained by direct numerical integration.
We can confirm that the approximate analytical estimate of the asymptotic phase given by Eq.~(\ref{eq:34}) gives reasonable agreement with the direct numerical results even though the system state is considerably far from the $\phi$-$R$ plane.

\subsection{Effect of a periodic force on the amplitude}

In this subsection, we consider the effect of a periodic external force of moderate intensity with small frequency detuning.
In particular, we focus on the average effect of the periodic force on the amplitude $R$ in the phase-locked state, which cannot be analyzed without the amplitude equation.

Since $g_{02} = 0$ in Eq.~(\ref{eq:28-1}), the $\psi$-nullcline on which $\dot{\psi} = 0$ is obtained from the averaged equation~(\ref{eq:psi_expand}) as
\begin{equation}
\psi=\frac{T}{2\pi}\arcsin\left[-\frac{1+Ra_{0}}{g_{01} G_{0}}\left(\Delta\omega+\frac{1}{1+Ra_{0}}\left(a_{2}R^{2}+a_{3}R^{3}\right)\right)\right]
\label{eq:34-1}
\end{equation}
when the argument of $\arcsin$ is in the range $[-1, 1]$.
By substituting Eq.~(\ref{eq:34-1}) into Eq.~(\ref{eq:R_expand}), we can obtain the fixed points of the averaged amplitude dynamics satisfying 
$\dot{R} = F_{s}\left(R,G_{0},\Delta\omega\right)=0$,
where $F_{s}$ represents the right-hand side of Eq.~(\ref{eq:R_expand}).
The effect of the intensity of the periodic force $G_{0}$ and the detuning $\Delta\omega$ on the stationary amplitude $R$ of the oscillation in the steady state can be evaluated from the partial derivatives of $F_{s}\left(R,G_{0},\Delta\omega\right)$
by the implicit function theorem.

\begin{figure}[th]
\includegraphics{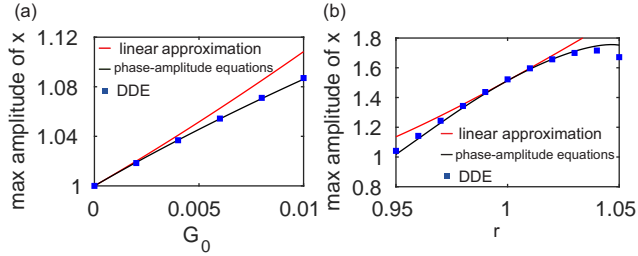}
\caption{Maximum amplitude of the DDE~(\ref{X}) subjected to a periodic input.
(a) Dependence of maximum amplitude on $G_{0}$ at $r=1$.
(b) Dependence of maximum amplitude on $r$
at $G_{0}=0.1$. Blue points show numerical results obtained by direct numerical integration of the original DDE~(\ref{X}). The red lines are analytical predictions by the linear approximation, while black lines are numerical solutions of the phase-amplitude equations~(\ref{eq:psi_expand}) and (\ref{eq:R_expand}).
}
\label{fig3-1-1}
\end{figure}

Figure~\ref{fig3-1-1} shows the predicted amplitude of the oscillation. The dependence of the amplitude on $G_{0}$ at $r = 1$ is plotted in Fig.~\ref{fig3-1-1}(a), where the stationary amplitude obtained from the averaged phase-amplitude equations~(\ref{eq:psi_expand}) and (\ref{eq:R_expand})
are compared with the linear approximation of the stationary amplitude with a slope ${\partial R} / {\partial G_{0}}\mid_{R=0,G_{0}=0,\Delta\omega=0}=18.2$.
Similarly, Fig.~\ref{fig3-1-1} (b) shows the dependence of the amplitude 
on $r$ at $G_{0}=0.1$, where the result of the phase-amplitude equations are
compared with linear approximation of the amplitude with a slope ${\partial R} / {\partial r}\mid_{R=0.72,G_{0}=0.1,\Delta\omega=0}=9.91$.
%
We can confirm that the linear approximation appropriately predicts the changes in the stationary amplitude of the delay-induced oscillator subjected to a non-weak external periodic force when it is slightly modulated. 
Moreover, 
the nonlinear phase-amplitude equations can predict
the amplitude more precisely than the linear approximation in the given parameter range. 

\subsection{Bistable response of delay-induced oscillation to a periodic force}

\begin{figure}[th]
\begin{centering}
\includegraphics{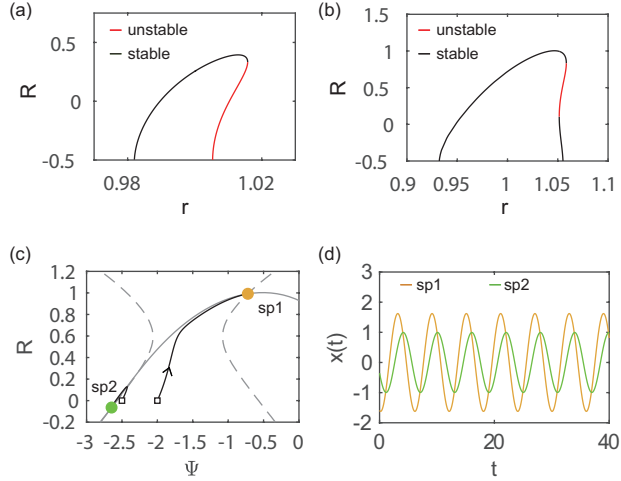}
\par\end{centering}
\caption{ (a) Stable and unstable fixed points
of the amplitude $R$ plotted against the frequency detuning $r$ at $G_{0}=0.02$.
(b) Stable and unstable fixed points at $G_{0}=0.1$. A bistable region exists near $r=1.052$.
(c) Nullclines and stable fixed points on the ($\psi$, $R$)-plane
at $r=1.052$ and $G_{0}=0.1$.
The gray broken lines show the nullclines satisfying $\dot{\psi}=0$ and the gray solid line shows $\dot{R}=0$. The orange and green dots show the stable fixed points (sp1 and sp2) and black lines show the trajectories started from $(-2, 0)$ and $(-2.5, 0)$. Sp1 is located at $(-0.722, 0.992)$
and sp2 is at $(-2.647, -0.064)$.
(d) Time course of Eq.~(\ref{X})
with $G(x,t)=0.1\sin\left(1.052t\right)$. The orange line corresponds
to the sp1, while the green line corresponds to the sp2 in panel (c). }
\label{fig3}
\end{figure}

In this subsection, we demonstrate that the present model can exhibit a nontrivial bistable response to a periodic force by a bifurcation analysis of Eq.~(\ref{eq:psi_expand}) and Eq.~(\ref{eq:R_expand}).
Such a phenomenon results from higher-order amplitude effects and cannot be described by the phase-only equation nor the lowest-order phase-amplitude equations. Using XPP-AUTO~\cite{XPP}, we numerically find stationary solutions in the range $R>-0.5$ where the inverse $1/(1+Ra_{0})$ exists (note that $a_{0} = 1.8418$).
Depending on the parameters $G_0$ and $r$, we observe quantitatively different behaviors of the system state as shown in Fig.~\ref{fig3}.

Figures~\ref{fig3} (a) and (b) show the stable and unstable fixed points
on the ($R$, $r$)-plane at two different values of $G_{0}$.
The system is always monostable when $G_{0}=0.02$, while a bistable region where $R$ can take two stable fixed points is found around $r=1.052$ when $G_{0}=0.1$.
Thus, it is expected that DDE (\ref{X}) with $G(x,t)=0.1\sin\left(1.052t\right)$ shows
bistable dynamics. 
Figure~\ref{fig3} (c) shows the nullclines and stable fixed points on the $\psi$-$R$ plane at $r=1.052$ and $G_{0}=0.1$.
The two crosses show the stable fixed points at $(-0.722, 0.992)$ and $(-2.647, -0.064)$,
and the two black lines show the trajectories started from $(-2, 0)$ and $(-2.5, 0)$.
These predictions from the reduced phase-amplitude equations can be confirmed in Fig.~\ref{fig3} (d), which shows the results of direct numerical integration of DDE (\ref{X}) with $G(x,t)=0.1\sin\left(1.052t\right)$.
We can clearly observe the bistable dynamics of the oscillator caused by moderately strong periodic forcing.

\section{Gene-regulatory oscillator}

In this section, as a more complex, biologically-motivated example of DDEs, we investigate a model of gene regulation~\cite{Dfiremother} under a periodic sinusoidal force given by 
\textbf{\textsl{\textcolor{orange}{{} }}}
\begin{equation}
\frac{dx(t)}{dt}=\frac{\alpha C_{0}^{2}}{[C_{0}+x(t-\tau)]^{2}}-\frac{\gamma x(t)}{R_{0}+x(t)}-\beta x(t) +G(t),\label{nonlinear}
\end{equation}
where $x(t) \in {\mathbb R}$ is the state variable representing protein concentration and $\alpha, \beta, \gamma, C_0, R_0$, and the delay time $\tau$ are real parameters. 
The first term of the right-hand side represents protein synthesis with time delay for transcription and translation, while the second and the third terms represent degradation and dilution of the protein, respectively.  
Following the previous research~\cite{Dfiremother}, we set  $\beta=0.1$, $C_{0}=10$,  and $\tau=1$. The external periodic force is $G(t)=G_{0}\sin\left(\frac{2\pi}{T} r t \right)$ with  intensity $G_0$ and frequency mismatch $r$. 
We set the rate constant of synthesis as $\alpha=100$, degradation as $\gamma=100$, and Michaelis constant of degradation as $R_{0}=10$ so that the system exhibits a slow convergence to a limit cycle orbit and the effect of the amplitude dynamics can be clearly observed. 

\begin{figure}[th]
\begin{centering}
\includegraphics{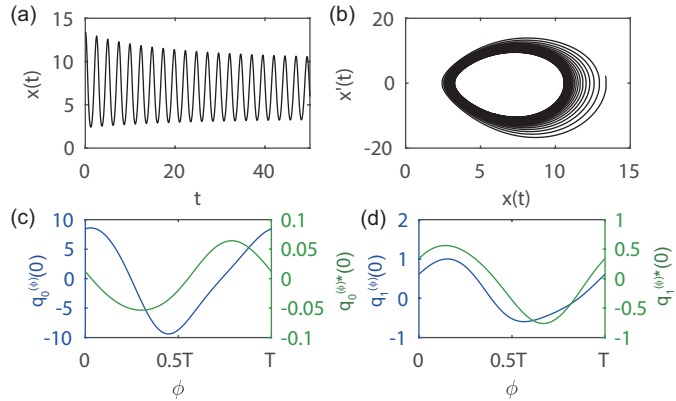}
\par\end{centering}
\caption{(a) Time course of the gene-regulatory oscillator Eq.~(\ref{nonlinear}) without perturbation ($G=0$), showing a slow
convergence to the limit cycle.
(b) Trajectory of the system state projected onto the $(x,dx/dt)$-plane.
(c) Floquet and adjoint eigenfunctions $q_{0}^{(\phi)}(0)$ and $q_{0}^{(\phi)^*}(0)$ associated with $\lambda_{0}=0$. (d) Floquet and adjoint eigenfunctions $q_{1}^{(\phi)}(0)$ and $q_{1}^{(\phi)^*}(0)$  associated with $\lambda_{1}$.
}
\label{fig3-1}
\end{figure}

This system has a stable limit cycle with a period $T=2.46$, which can be obtained only numerically.
Figures~\ref{fig3-1} shows the system state $x^{(t)}$ converging toward the limit-cycle attractor; Fig.~5(a) plots the time course of $x(t)$ as a function of $t$ and Fig.~5(b) shows the system trajectory projected on the $(x,dx/dt)$-plane.
The time constant of the relaxation to the limit cycle is much larger than the period of the oscillation as can be seen from the figures.

For this model, the $T$-periodic linear operator $\hat{L}$ is given by Eq.~(\ref{eq. hatL}) with
\begin{equation}
{\bf \bar{\Omega}}^{(t)}(\sigma)=\delta(\sigma)\left\{ -\beta-\frac{\gamma R_{0}}{\left(R_{0}+x_{0}(t)\right)^{2}}\right\} +\delta(\sigma+\tau)\left\{ \frac{-2\alpha C_{0}^{2}}{\left(C_{0}+x_{0}(t-\tau)\right)^{3}}\right\} .
\end{equation}
Figures~\ref{fig3-1} (c) and~\ref{fig3-1}(d) show the first two eigenfunctions and adjoint eigenfunctions of $\hat{L}$ obtained by the extended adjoint method~\cite{origin}, respectively.
The second largest Floquet exponent is $\lambda_{1} = -0.0255$ in this case.
From these eigenfunctions, the phase-amplitude equations~(\ref{eq:psi_expand}) and~(\ref{eq:R_expand}) under the sinusoidal force can be obtained, where the coefficients are given by
$a_0=0.330$, $a_2=-5.33 \times 10^{-4}$, $a_3=1.13 \times 10^{-4}$, $g_{01}=-0.0296$, $g_{02}=-7.57 \times 10^{-3}$, $b_0=-2.48$, $b_2=-4.74  \times 10^{-3}$, $b_3=1.15  \times 10^{-3}$,
$g_{11}=-0.176$, and $g_{12}=0.282$.

\begin{figure}[th]
\begin{centering}
\includegraphics{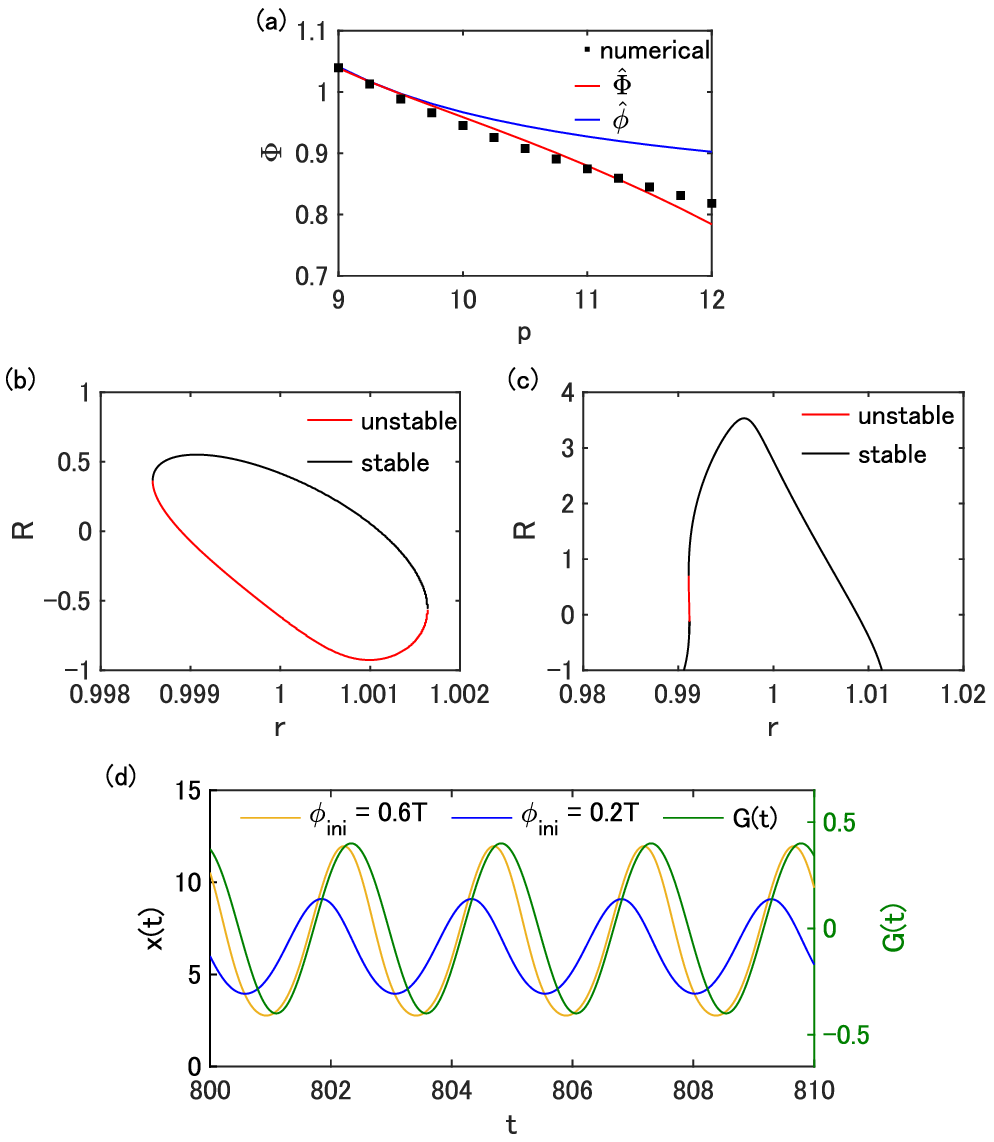}
\par\end{centering}
\caption{(a) Asymptotic phase values of initial functions $x^{(t=0)} (\sigma)\equiv p$ far from the limit cycle.
The black points indicate the asymptotic phase obtained by direct numerical integration, the blue line indicates analytical estimation of the phase $\hat{\phi}$,
and the red line indicates analytical estimation of the asymptotic phase $\hat{\Phi}$.
(b) Stable and unstable fixed points
plotted with respect to $r$ at $G_{0}=0.05$.
(c) Fixed points at $G_{0}=0.4$. A bistable region exists around $r=0.9911$.
(d) Time course of the DDE~(\ref{nonlinear}) with $G_0=0.4$ and $r=0.9911$. The red line shows the result for the initial condition {$x^{(t=0)} (\sigma)=X_0^{(\phi_{{\rm ini}}=0.6T)} (\sigma)$, while the blue shows the result for $x^{(t=0)} (\sigma)=X_0^{(\phi_{{\rm ini}}=0.2T)} (\sigma)$ ($-\tau \leq \sigma \leq 0$). The green line represents the external force.}
}
\label{fig5}
\end{figure}

We first evaluate the validity of the approximate expression of the asymptotic phase in the absence of the external force ($G=0$). We take the initial condition as a constant function, $x^{(t=0)}(\sigma)=p$, and evaluate the asymptotic phase by Eq.~(\ref{eq:R3}) and by direct numerical integration of the DDE.
Figure~\ref{fig5} (a) shows the phase ${\phi}$, the asymptotic phase ${\Phi}$ estimated by using Eq.~(\ref{eq:R3}), and the asymptotic phase $\Phi$ evaluated by direct numerical integration of the DDE.
It can be seen that the approximate analytical result reproduces the result of direct numerical measurement of the asymptotic phase.

We next consider how the gene-regulatory oscillator behaves when it is subjected to a periodic external force.
We conduct bifurcation analysis for different values of $G_0$ and $r$ in the same way as that for Eq.~(\ref{X}) using XPP-AUTO.
When the external periodic force is weak ($G_0=0.05$) and the frequency mismatch is small enough, the system is synchronized to the periodic force with a single stable amplitude as shown in Fig.~\ref{fig5}(b), namely, the amplitude response is monostable.
When we apply a stronger force, $G_{0}=0.4$, the region of synchronization becomes wider. The amplitude of synchronized oscillations is positive when the frequency mismatch is small, whereas the amplitude is negative when the mismatch is large.
Moreover, there exists a bistable region around $r=0.99$ as shown in Fig.~\ref{fig5}(c), where $R$ can take either of two stable values, similar to the previous simpler model with a cubic nonlinearity described by Eq.~(\ref{X}).

Figure~\ref{fig5} (d) shows two time courses of DDE (\ref{nonlinear}) with $G_{0}=0.4$ and $r=0.9911$ with different initial conditions. In this case, a small-amplitude out-of-phase oscillation emerges in addition to the large-amplitude oscillation that
exists in a wider range of $r$. Both types of oscillations are stable.
In the video in the Supplemental Material~\cite{supp}, the slow convergence of the system state to either of these two oscillatory states are visualized by projecting the system state onto the $(x,dx/dt)$-plane. 
It is noteworthy that the frequency mismatch required for this bistable dynamics is very small (less than 1 \%) in this model. 

\section{Summary}

In this study, we have developed a general mathematical framework for reducing delay-induced limit-cycle oscillators described by DDEs into a set of nonlinear phase-amplitude equations on the basis of the Floquet theory.
By projecting the original equation onto the reduced phase space spanned by the first two Floquet eigenfunctions, we derived a set of nonlinear phase-amplitude equations.
We proposed an extended adjoint method for DDEs to numerically calculate the Floquet eigenfunctions and their adjoint eigenfunctions.
We also developed a method to estimate the asymptotic phase of the system states in a neighborhood of the limit cycle from the phase and amplitude defined by the Floquet eigenfunctions.
The validity of the framework has been confirmed by analyzing two models of delay-induced oscillations.
In the present framework, the derivation of the reduced equations requires only the calculation of the first two Floquet and adjoint eigenfunctions.
Therefore, the reduction is practically manageable even though the dynamical system to be reduced is an infinite-dimensional DDE.

Despite the simplicity, the resulting reduced equations convey richer information than simply linearizing the system state around the periodic orbit.
To illustrate this, we first studied an analytically tractable DDE with a cubic nonlinearity. We derived an approximate expression of the nonlinear asymptotic phase in terms of the phase and amplitude and verified its validity using direct numerical integration of the original system.
Moreover, we revealed nontrivial bistable synchronization of the system with a periodic external forcing, where the amplitude can take two different stable values depending on the initial condition, which cannot be analyzed within the conventional phase-only or the lowest-order phase-amplitude equations. 
We also analyzed a model of gene-regulatory oscillator and showed that the reduced phase-amplitude equations also enabled us to capture the nontrivial bistable synchronization with a non-weak periodic force.

The result for the gene-regulatory oscillator provides analytical insights into how the weak attraction of the limit cycle and nonlinear interactions between the phase and amplitude can alter the synchronization dynamics of gene regulatory systems for circadian oscillations.
For example, it is known that, in the case of ASPS, out-of-phase (phase-advanced) synchronized oscillation with the day-and-night lights is stabilized in a similar manner to that is shown in Fig.~\ref{fig5}(d) of the second model. It has also been reported that the free-running period of circadian oscillation in ASPS patients is shorter than 24 h~\cite{Jones}, and the temporal therapy (phase advance chronotherapy) can alter the out-of-phase synchronization into in-phase synchronization
~\cite{Moldofsky}.
Our theoretical results imply that weak attraction of the limit cycle and nonlinear interactions between the phase and amplitude could induce small-amplitude oscillations and bistability of the out-of phase and in-phase synchronized states.
If this is the case, the rate of attraction of the system state to the limit cycle, the Floquet exponent $\lambda_1$ in our study, could be used as another effective index to understand circadian rhythm disorders
in addition to conventional indices like the free-running periods and amplitudes of oscillation~\cite{Lema+, Thorpy+, Jones, Ukai+}.
Thus, the phase-amplitude analysis of delay-induced oscillations developed in this study can shed new light on the complex biological rhythms.

There are many other examples of natural and artificial systems that exhibit
complex oscillations due to the effect of time delay~\cite{Glass+,Lewis+, Dfiremother, Brent+, Peterka+,Soriano,Kalmar+,Szydlowski+}. For example, breathing of chronic heart failure patients is a typical example of such natural systems~\cite{Glass+}. The present study would provide further insights into nontrivial breathing dynamics. An example of artificial systems is the Mackey-Glass electrical circuit~\cite{MGC} that can be modeled by a DDE, for which the present theory is readily applicable to analyze the synchronization dynamics.
The present framework for reducing such time-delayed systems to a set of
nonlinear phase-amplitude equations can be useful as a general analytical method to
elucidate the origin of complex synchronization properties
under the effect of non-weak perturbations or fluctuations.
Further investigation on the nonlinear phase-amplitude equations would provide us with
more insight into the synchronization dynamics in time-delayed systems.

\section{Acknowledgments}
We thank Yuki Shimono for helpful advice on computation of the eigenfunctions
and G. Bard Ermentrout for fruitful discussion. 
This study is supported in part by JSPS KAKENHI (18H04122), The Asahi Glass Foundation, and JST PRESTO (JPMJPR14E2) to KK, and JSPS KAKENHI (JP16K13847, JP17H03279, 18K03471, and JP18H03287) and JST CREST (JPMJCR1913) to HN.

\appendix

\section{Derivation of the nonlinear phase-amplitude equations\label{app. pae}}

In this section, details of the derivation of the phase-amplitude equations are presented.
We first define a phase $\phi\in\left[0,T\right)$ along the unperturbed
limit cycle of Eq.~(\ref{2}), and represent the $T$-periodic eigenfunctions $q_{j}^{(t)}$
as functions of the phase $\phi\left(t\right)$ as $q_{j}^{(\phi)} $,
where $\phi\left(t\right)=t \: (\mathrm{mod} \: T)$.
Because we assume that the functional components associated with the eigenvalues $\lambda_i \: (i \ge 2)$ decay quickly, we approximate the system state $X^{(t)}$ as $X^{(t)}({\sigma})\simeq X_{0}^{(\phi)}({\sigma})+R q_{1}^{(\phi)}(\sigma)$ ($-\tau\leq\sigma\leq0$), where $X_0^{(\phi)}$ is the system state with phase $\phi$ on the limit cycle and $R$ is the amplitude of the eigencomponent corresponding to $\lambda_1$.
Substitution of this approximation into the functional differential equation (\ref{2}) yields
\begin{eqnarray}
&&
\left[\frac{d}{d\phi}X_{0}^{(\phi)}(\sigma)+R\frac{d}{d\phi}q_{1}^{(\phi)}\left(\sigma\right)\right]\dot{\phi}+q_{1}^{(\phi)}\left(\sigma\right)\dot{R}
\cr
&&=
\begin{cases}
\frac{d}{d\sigma}X_{0}^{(\phi)}(\sigma)+R\frac{d}{d\sigma}q_{1}^{(\phi)}\left(\sigma\right), &(-\tau\leq\sigma<0)\\
\begin{split}
& \mathcal{N}(X_{0}^{(\phi)} (\cdot) )+R\int_{-\tau}^{0}d\sigma'{\bf \bar{\Omega}}^{(\phi)}(\sigma')q_{1}^{(\phi)}(\sigma')
+F_{\mathrm{nl}}\left(\phi,R\right)+G\left(\phi,R,t\right),
\end{split} &(\sigma=0)
\end{cases}\label{eq:10}
\end{eqnarray}
where
\begin{equation}
F_{\mathrm{nl}}\left(\phi,R\right)=\mathcal{N}\left(X_{0}^{(\phi)} (\cdot) +Rq_{1}^{(\phi)}\left(\cdot\right)   \right)-\mathcal{N}\left(X_{0}^{(\phi)} (\cdot) \right)-R\int_{-\tau}^{0}d\sigma'{\bf \bar{\Omega}}^{(\phi)}(\sigma')q_{1}^{(\phi)}\left(\sigma'\right).
\end{equation}

To derive the phase equation, we project both sides of Eq.~(\ref{eq:10})
onto the eigenfunction $q_{0}^{(\phi)}$.
Using the relations
\begin{align}
\frac{d}{d\phi}X_{0}^{(\phi)}(\sigma)=q_{0}^{(\phi)}\left(\sigma\right),
\end{align}
\begin{align}
\frac{d}{d\phi}q_{1}^{(\phi)}\left(\sigma\right)=-\lambda_{1}q_{1}^{(\phi)}\left(\sigma\right)+L^{(\phi)}\left(q_{1}^{(\phi)}\right)\left(\sigma\right),
\end{align}
and
\begin{align}
\mathcal{N}\left(X_{0}^{(\phi)} (\cdot) \right)=q_{0}^{(\phi)}\left(0\right),
\end{align}
which follows from the definition $q_{0}^{(\phi)}\left(0\right) = dX_{0}/dt|_{t}$,
we obtain
\begin{eqnarray}
\left[1+R \left\langle q_{0}^{(\phi)*}, L^{(\phi)} \left(q_{1}^{(\phi)}\right);\phi\right\rangle \right]\dot{\phi} &=& 1+R\left\langle q_{0}^{(\phi)*},L^{(\phi)}\left(q_{1}^{(\phi)}\right);\phi\right\rangle \nonumber \\
&& + q_{0}^{(\phi)*}\left(0\right)\cdot\left(F_{\mathrm{nl}}\left(\phi,R \right)+G\left(\phi,R,t\right)\right).
\end{eqnarray}
The phase equation is thus given by
\begin{eqnarray}
\dot{\phi}&=&1+\frac{q_{0}^{(\phi)*}\left(0\right)\cdot\left(F_{\mathrm{nl}}\left(\phi,R\right)+G\left(\phi,R,t\right)\right) }{1+R\left\langle q_{0}^{(\phi)*},L^{(\phi)}\left(q_{1}^{(\phi)}\right);\phi\right\rangle } \nonumber\\
&=&1
+ q_{0}^{(\phi)*}\left(0\right)\cdot\left(F_{\mathrm{nl}}\left(\phi,R\right)+G\left(\phi,R,t\right)\right)  \nonumber\\
&&-\frac{R\left\langle q_{0}^{(\phi)*},L^{(\phi)}\left(q_{1}^{(\phi)}\right);\phi\right\rangle}{1+R\left\langle q_{0}^{(\phi)*},L^{(\phi)}\left(q_{1}^{(\phi)}\right);\phi\right\rangle }
q_{0}^{(\phi)*}\left(0\right)\cdot\left(F_{\mathrm{nl}}\left(\phi,R\right)+G\left(\phi,R,t\right)\right).
\label{eq:13}
\end{eqnarray}

Similarly, by projecting both sides of Eq.~(\ref{eq:10}) onto the eigenfunction $q_{1}^{(\theta)}$, we obtain
\begin{eqnarray}
\left[R\left\langle q_{1}^{(\phi)*}, \left(-\lambda_{1}q_{1}^{(\phi)}+L^{(\phi)}\left(q_{1}^{(\phi)}\right)\right);\phi\right\rangle \right]\dot{\phi}+\dot{R} &=& R\left\langle q_{1}^{(\phi)*}, L^{(\phi)} \left(q_{1}^{(\phi)}\right);\phi\right\rangle \nonumber \\
&& + q_{1}^{(\phi)*}\left(0\right)\cdot\left(F_{\mathrm{nl}}\left(\phi,R\right)+G\left(\phi,R,t\right)\right).
\label{eq:15}
\end{eqnarray}
By substituting Eq.~(\ref{eq:13}) into Eq.~(\ref{eq:15}), the amplitude equation is derived as
\begin{eqnarray}
\dot{R}&=&\lambda_{1}R
+q_{1}^{(\phi)*}\left(0\right)\cdot\left(F_{\mathrm{nl}}\left(\phi,R\right)+G\left(\phi,R,t\right)\right)
\cr
&-&\frac{R\left[\left\langle q_{1}^{(\phi)*}, L^{(\phi)}\left(q_{1}^{(\phi)}\right);\phi\right\rangle -\lambda_{1} \right]}{1+R\left\langle q_{0}^{(\phi)*},L^{(\phi)}\left(q_{1}^{(\phi)}\right);\phi\right\rangle} q_{0}^{(\phi)*}\left(0\right)\cdot\left(F_{\mathrm{nl}}\left(\phi,R\right)+G\left(\phi,R,t\right)\right)
 \label{eq:16}
\end{eqnarray}

\section{Coefficients of the phase-amplitude equations\label{app:coefficients}}

The expressions for the individual expansion coefficients in the phase and amplitude equations~(\ref{eq:psi_expand}) and~(\ref{eq:R_expand}) are as follows.

\begin{align}
a_{0}=\frac{1}{T}\int_{0}^{T}\left\langle q_{0}^{(\theta)*},L^{(\theta)}\left(q_{1}^{(\theta)}\right);\theta\right\rangle d\theta, 
\end{align}
\begin{align}
a_{2}=\frac{1}{T}\int_{0}^{T}q_{0}^{(\theta)*}\left(0\right)\cdot F_{\mathrm{nl},2}\left(\theta\right)d\theta,
\end{align}
\begin{align}
a_{3}=\frac{1}{T}\int_{0}^{T}q_{0}^{(\theta)*}\left(0\right)\cdot F_{\mathrm{nl},3}\left(\theta\right)d\theta,
\end{align}
\begin{align}
g_0(\psi) = \frac{1}{T} \int_{0}^{T} q_{0}^{(\theta)*}\left(0\right)\cdot G \left( \frac{\theta - \psi}{r} \right) d\theta,
\end{align}
\begin{align}
b_{0}=\frac{1}{T}\int_{0}^{T}\left\langle q_{1}^{(\theta)*},L^{(\theta)}\left(q_{1}^{(\theta)}\right);\theta\right\rangle d\theta, 
\end{align}
\begin{align}
b_{2}=\frac{1}{T}\int_{0}^{T}q_{1}^{(\theta)*}\left(0\right)\cdot F_{\mathrm{nl},2}\left(\theta\right)d\theta,
\end{align}
\begin{align}
b_{3}=\frac{1}{T}\int_{0}^{T}q_{1}^{(\theta)*}\left(0\right)\cdot F_{\mathrm{nl},3}\left(\theta\right)d\theta, 
\end{align}
and
\begin{align}
g_1(\psi) = \frac{1}{T} \int_{0}^{T} q_{1}^{(\theta)*}\left(0\right)\cdot G \left( \frac{\theta - \psi}{r} \right) d\theta.
\end{align}

\section{Supplementary video}

This supplementary video shows how the oscillators converge to either of the two stable oscillatory states.
We numerically integrated the DDE~(\ref{nonlinear}) of $15$ gene-regulatory oscillators
subjected to a common sinusoidal external periodic force from several initial conditions from $t=0$ to  $t=10^3$.
The parameters of the external force are $G_{0}=0.4$ and $r=0.9911$.
The initial amplitude and phase of each oscillator is $R_k=0$ and $\phi_k = \frac{k}{15}T~(k=1,2,...,15)$.

In panel (a) of the video, the states of the oscillators projected onto the $(x,dx/dt)$-plane are plotted.
In panel (b) of the video, the time courses from two representative initial conditions are shown, where the magenta line is for $\phi_{\mathrm{ini}}=\frac{7}{15}T$ and the blue one is for $\phi_{\mathrm{ini}}=\frac{2}{15}T$. 

\clearpage



\begin{thebibliography}{10}
\bibitem{bilinear}
The dual space of $C_0$ is usually defined as a vector space consisting of all bounded linear functionals with domain $C_0$.
However, it is difficult to work directly with this space, because, in general,
the adjoint system on this space is not described by a differential equation,
i.e., the adjoint semigroup is not strongly continuous~\cite[Ch.~2]{Diekmann}.
Hale introduced a bilinear form, under which the adjoint semigroup
on the associated adjoint space is strongly continuous.
In the celebrated work of Hale and Lunel, the convenient system is called
not the adjoint system but the transposed system to distinguish between them.
The transposed system can be viewed as the "adjoint" system
since their eigenfunctions are closely related~\cite[Chs.~7 and 8]{Hale}.
Hence, we call the transposed system as the adjoint system in this paper as in the literature~\cite{Simmendinger}.
\bibitem{footnotesec3c}
The frequency of this subtraction procedure may not necessarily be once per time $T$. Generally speaking, the faster the $q_0$ component grows relative to $q_1$, the more frequent subtraction would be required. In this study, at every 10 oscillation periods, whether $Y^{(t)}$ decays exponentially or not is judged by measuring 10 peaks of $Y^{(nT)}$. If this is the case, the Floquet eigenvalue $\lambda_{1}$ is evaluated using the peaks (Fig.~\ref{figs}(c)). The waveform of $q_{1}^{(t)}(\sigma)$ is obtained by averaging $q_{1}^{\left(t\right)}\left(\sigma\right)=e^{-\lambda_{1}t}\cdot Y^{(t)}(\sigma)$ over the 10 oscillation periods, where exponential decay is compensated (Fig.~\ref{figs}(d), (e)).
\bibitem{lambda_complex}
The eigenspace associated with the complex conjugate eigenvalues are
numerically obtained by the consecutive subtraction approach proposed in 
Section II B, but some care is required when evaluating the pair of complex conjugate eigenfunctions.
Ding and Cvitanovi\'{c} have developed an algorithm to calculate the
complex Floquet eigenvalues and eigenfunctions~\cite{Ding}.
By introducing a moving coordinate frame spanned by the complex eigenfunctions as in Section~\ref{sec. pa}, a reduced ODE form may be obtained even in such a situation.
Though not generic, Floquet eigenvalues with multiplicities can also arise,
whose treatment is beyond the scope of this paper.
We here only mention that, in some mathematical literature,
spectral projections of the time-$T$ fundamental solution of
linear periodic DDEs onto generalized eigenspaces associated with multiple eigenvalues
has been discussed on the basis of the Dunford integral using residue calculus~\cite{Diekmann,Frasson}.
\bibitem{comment1} It is of note that the phase remains zero regardless
of the value of $p$ when we naively evaluate it based on an argument introduced on a plane $(x,\dot{x})$ as $\tan^{-1}(x/\dot{x})$.
\bibitem{origin}
The origin of the phase is set at $t$ that satisfies $x_0(t)=\langle x_0(t)\rangle$ with $dx_0(t)/dt>0$.
\end{thebibliography}
\end{document}